\documentclass{aa}  
\usepackage{graphicx}
\usepackage{mathtools}
\usepackage{amsmath}
\usepackage{fancyref}
\usepackage{longtable}
\usepackage{rotating}
\usepackage{pdflscape}
\usepackage[breaklinks=false, colorlinks=true, citecolor=blue,urlcolor=blue,filecolor=blue, linkcolor=blue]{hyperref}
\usepackage{txfonts}
\usepackage{natbib}
\def\myurl{\hfil\penalty 100 \hfilneg \hbox}

\begin{document}

   \title{Search for associations containing young stars (SACY)}

   \subtitle{V. Is multiplicity universal? Tight multiple systems\thanks{Based on observations obtained using the instruments FEROS at La Silla (ESO 1.5m and MPG/ESO 2.2m) UVES at VLT (ESO 8m)}$^{,}$\thanks{Appendices C and D are available in electronic form at 
\myurl{\url{http://www.aanda.org}}}$^{,}$\thanks{Tables~\ref{table:av_values_all_sources}, 
\ref{table:sb2_vr_values}, 
and \ref{tab:member_list} 
are available in electronic form at the CDS via anonymous ftp to \nolinkurl{cdsarc.u-strasbg.fr (130.79.128.5)} or via \url{http://cdsarc.u-strasbg.fr/viz-bin/qcat?J/A+A/568/A26}}}

   \author{P. Elliott
          \inst{1, 2}\and
          A. Bayo
	  \inst{1,3,4}\and
	  C. H. F. Melo
	  \inst{1}\and
	  C. A. O. Torres
	  \inst{5}\and
	  M. Sterzik
	  \inst{1}\and
	  G. R. Quast
	  \inst{5}
          }
   \institute{European Southern Observatory, Alonso de Cordova 3107, Vitacura Casilla 19001, Santiago 19, Chile
              \\
              \email{pelliott@eso.org}
         \and
             School of Physics, University of Exeter, Stocker Road, Exeter, EX4 4QL
	\and 
             Max Planck Institut f\"ur Astronomie, K\"onigstuhl 17, 69117, Heidelberg, Germany
	\and 
            Departamento de F\'isica y Astronom\'ia, Facultad de Ciencias, Universidad de Valpara\'iso, Av. Gran       Breta\~na 1111, 5030 Casilla, Valpara\'iso, Chile
	 \and
	    Laborat\'orio Nacional de Astrof\'isica/ MCT, Rua Estados Unidos 154, 37504-364 Itajub\'a (MG), Brazil
             }

   \date{Received 21 March 2014 / Accepted 5 June 2014}

 
  \abstract
   {Dynamically undisrupted, young populations of stars are crucial in studying the role of multiplicity in relation to star formation.  Loose nearby associations provide us with a great sample of close ($<$150 \,pc) pre-main sequence (PMS) stars across the very important age range ($\approx$5-70\,Myr) to conduct such research.}
   {We characterize the short period multiplicity fraction of the search for associations containing young stars (SACY) sample, accounting for any identifiable bias in our techniques and present the role of multiplicity fractions of the SACY sample in the context of star formation.}
   {Using the cross-correlation technique we identified double-lined and triple-lined spectroscopic systems (SB2/SB3s), in addition to this we computed radial velocity (RV) 
values for our subsample of SACY targets using several epochs of fiber-fed extended range optical spectrograph (FEROS) and ultraviolet and visual echelle spectrograph (UVES) data. These values were used to revise the membership of each association that was then combined with archival data to determine significant RV variations across different data epochs characteristic of multiplicity; single-lined multiple systems (SB1).}
   {We identified seven new multiple systems (SB1s: 5, SB2s: 2). We find no significant difference between the short period multiplicity fraction ($F_\mathrm{m}$) of the SACY sample and that of close star-forming regions ($\approx$1-2\,Myr) and the field ($F_\mathrm{m}\leq$10\%).  These are seen both as a function of age and as a function of primary mass, $M_1$, in the ranges $P$ [1:200\,day] and $M_2$ [0.08\,$M_{\odot}$-$M_1$], respectively. }
   {Our results are consistent with the picture of universal star formation, when compared to the field and close star-forming regions (SFRs).  We comment on the implications of the relationship between increasing multiplicity fraction with the primary mass within the close companion range in relation to star formation.}

 
   \maketitle
%

\section{Introduction}

Multiple stellar systems are abundant in our solar neighbourhood \citep{Duquennoy1991, Raghavan2010} and in our galaxy; they can be found in a variety environments, from young star-forming regions (SFRs) \citep{Ghez1997, Nguyen2012} to older, sparser populations such as globular clusters \citep{Sollima2007}.  These numerous multiple systems appear to be the default of nature and therefore studying them is studying a key ingredient of star formation and evolution.  

When studying the relevance of multiplicity in the context of star formation, two elements, mass and age, are key.  Different masses can provide information on the formation mechanism of the stellar systems \citep{Duquennoy1991, Fischer1992, Chini2012}.  The age of the system is related to the evolutionary state and history of dynamical effects \citep{Mason1998}. 
One must also consider the surrounding environment of the star; the density of stars increases the probability of dynamical interaction and perturbation whereby binary systems can be destroyed \citep{Kroupa1995, Parker2009}.


\begin{figure}[h]
\begin{center}
\includegraphics[width=0.45\textwidth]{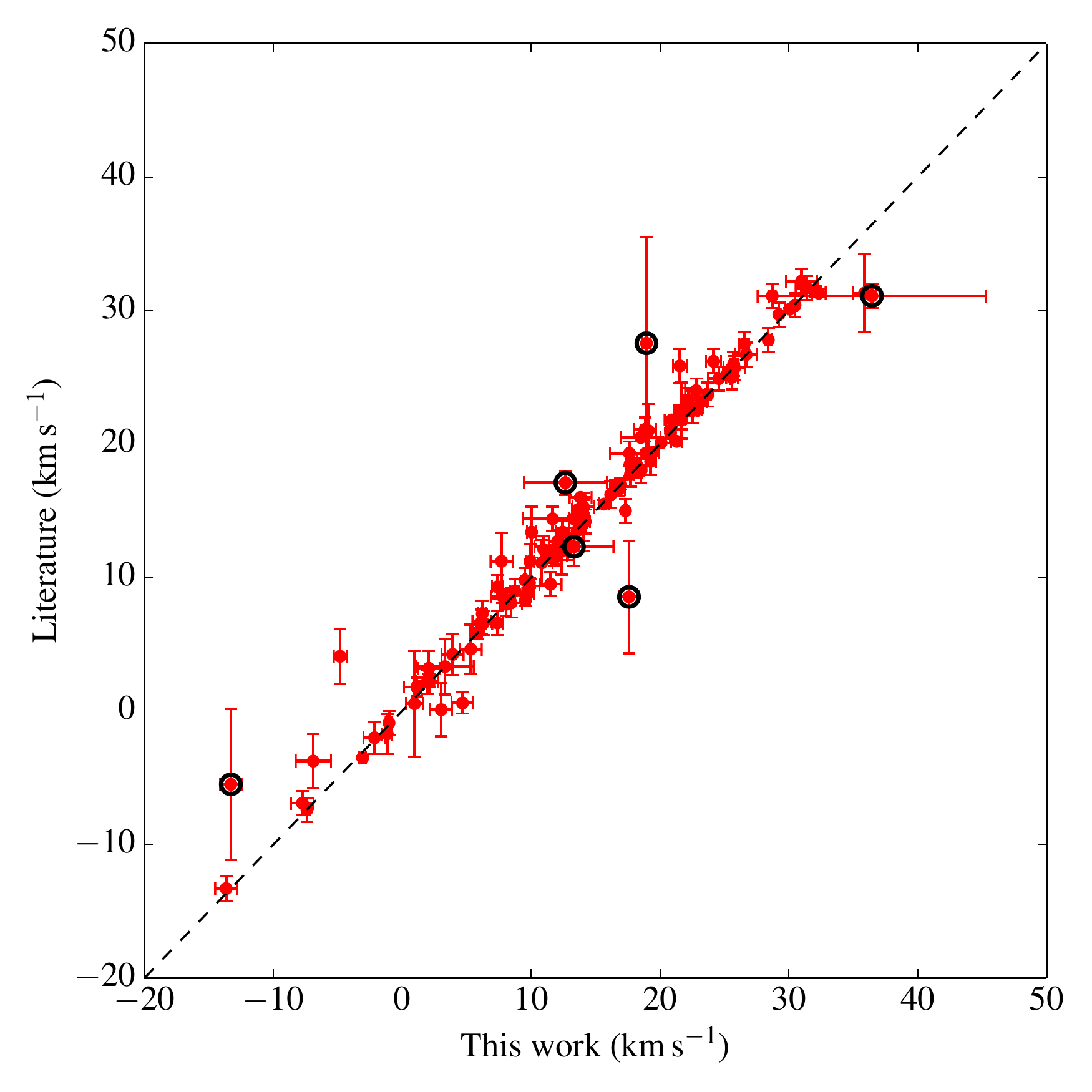}
\end{center}
\vspace{-0.5cm}
\caption{Average RV values determined in this work versus those from the literature, where values were available for comparison. Error bars represent the standard deviation from the mean value for each source.  If uncertainties were not quoted in the literature, the average uncertainty from all other values, 0.90\,km\,s$^{-1}$, was used. If only one data epoch was available, the overall standard deviation for the SACY sample, 0.89\,km\,s$^{-1}$, derived in this work was used to represent the uncertainty.  Black rings indicate identified multiple systems. A line representing a 1:1 relation has been drawn for base comparison.}
\label{fig:lit_comparison}
\end{figure}

The seminal work of \cite{Duquennoy1991} along with the more recent work of \cite{Raghavan2010} provide robust statistics of multiplicity frequency for solar-type stars in the field.  Combining results from studies of higher-mass stars \citep{Kouwenhoven2007, Chini2012} and lower-mass stars \citep{Henry1990, Fischer1992} the multiplicity frequency can be shown as a function of mass, as seen in Figure 12 of \cite{Raghavan2010}. 
%
%
The field stars used in such studies have been dynamically processed in different ways \citep{Parker2009}. In contrast, stars in associations are considered coeval with low average stellar densities and are much younger $\sim$1-100\,Myr.
%
%
We can therefore use these populations of stars as samples from which the effects of dynamical processing are very small and the observables are much closer to the direct outcome of star formation. 

%
%
%
%
The coeval populations that are usually studied comprise young nearby SFRs (1-5\,Myr), such as Taurus-Auriga, Ophiuchus-Scorpius, and Chamaeleon \citep{Melo2003, Ghez1993, King2012a}, across a wide range of orbital periods ($\sim$1-1000\,A.U.).  The results from these environments are difficult to compare due to variable extinction properties and large distances ($\ge$ 135\,pc).  However, some studies aim to do this \citep{King2012a, King2012b} using the limited overlapping parameter space between populations to produce robust multiplicity statistics.



In addition to SFRs, close associations, which are loose collections of stars sharing common kinematics, can offer populations that are close enough to the Sun to probe a large and continuous range of orbital periods. They also provide targets to study the transitional phase of disk dispersal.  The search for associations containing young stars (SACY) sample is comprised of nine close associations, as detailed in Table~\ref{tab:member_list}, with ages of $\sim$5-70\,Myr and distances of $<$150\,pc \citep{Torres2006, Torres2008}.  This paper is part of a series studying the SACY associations, concentrating on their multiplicity properties relating to star formation and disk evolution.  The multiplicity results presented in this paper are derived spectroscopically, probing the short period systems of the SACY sample and are compared to other populations, such as young close SFRs and the field.

{\tiny
\begin{table}
\caption{Summary of the properties of SACY associations studied in this work.  The age uncertainties provided represent the dispersion found in the literature. This dispersion is large; however, the relative aging performed with Li should not be affected.}
\begin{tabular}{p{1.59cm} p{0.5cm} p{0.8cm} p{0.4cm} p{0.9cm} p{0.5cm} p{0.5cm}}
\hline\hline\\
  \multicolumn{1}{l}{Association} &
  \multicolumn{1}{l}{Ass.} &
  \multicolumn{1}{l}{Dist.} &
  \multicolumn{1}{l}{Age} &
  \multicolumn{1}{l}{+ / -} &
  \multicolumn{1}{l}{N. of} &
  \multicolumn{1}{l}{N. of} \\
  \multicolumn{1}{l}{} &
  \multicolumn{1}{l}{ID} &
  \multicolumn{1}{l}{(pc)} &
  \multicolumn{1}{l}{(Myr)} &
  \multicolumn{1}{l}{(Myr)} &
  \multicolumn{1}{l}{Obj.} &
  \multicolumn{1}{l}{SBs} \\[1ex]
\hline\\
AB Doradus &              ABD & 34$\pm$26        & 70 &     30 / 20   &  53   & 5  \\
Argus &                       ARG & 106$\pm$51     & 40  &       10 / 30    & 33    & 0 \\
$\beta$-Pic &              BPC & 31$\pm$21        & 10  &      10 / 8    & 29   & 4\\
Carina &                       CAR & 85$\pm$35       & 30  &      5 / 20    & 22    & 3\\
Columba &                   COL & 82$\pm$30       & 30  &      40 / 15    & 31    &  1\\
$\epsilon$-Cha &        ECH & 108$\pm$9       & 6    &      1 / 0      & 20    &  2\\
Octans &                      OCT & 141$\pm$34     & 20  &      10 / 10    &  13   &  1\\
Tuc-Hor &                    THA & 48$\pm$7          & 30 &      0 / 20     & 33  &  0\\
TW-Hydrae &               TWA & 48$\pm$13       & 8     &     2 / 3      & 11    &  0\\[1ex]
\hline
\end{tabular}
\label{table:summary_associations}
\end{table}}



The paper is organised as follows. Section~\ref{sec:Observations and Data} details the observations and data used in this work.  Section~\ref{section:characterisation} describes the use of new radial velocity (RV) data to determine association membership and details the estimation of the mass distribution of the observed sample.  Section~\ref{sec:Identifying Multiple Systems} describes the methods employed to find multiple systems within our sample and the methods to account for biases.  Section~\ref{section:Results} presents the results and their significance in relation to star formation.  Finally in Section~\ref{sec:Conclusions and Further Work} the conclusions of this work on multiplicity and future prospects are presented.  Appendix~\ref{appendix:temp} shows the spectral type to effective temperature conversion. Appendix~\ref{sec:Individual Sources} notes individual sources; Appendix~\ref{appendix:vrad} details of the individual RV values, and Appendix~\ref{appendix:members} the targets considered as members in the SACY sample.







\section{Observations and data}
\label{sec:Observations and Data}

There are two main sources of high resolution spectroscopy available to us.\\

{First, observations performed with the fiber-fed extended range optical spectrograph (FEROS), a high resolution echelle spectrograph ($\lambda/\Delta\lambda\sim50,000$) at La Silla, Chile in two distinct periods: between January 1999 and September 2002 at the 1.5m/ESO telescope (ESO program ID: 67.C-0123) and after October 2002 at the 2.2m telescope (ESO program IDs: 072.C-0393, 077.C-0138).  These data were reduced and RV values were calculated for these targets as described in \cite{Torres2006}.  Out of the total SACY sample used in this work, 127 sources have existing RV values from FEROS data and multiplicity flags.\\

Second, observations performed with the ultraviolet and visual echelle spectrograph (UVES) ($\lambda/\Delta\lambda\sim40,000$ with 1$\arcsec$ slit)
 at Paranal, Chile with ESO program IDs: 088.C-0506(A) \& 089.C-0207(A) between October 2011 and September 2012. In these two programs each source had three separate observations using the standard set-up, a 1$\arcsec$ slit width, which covers the wavelength range 3250-6800\,$\AA$.  The separation in time between the acquisition of each epoch of data for one source ranges from 1 day to $\sim$ 1 month; the simulations described in Section~\ref{subsec: bias} help to characterize possible bias from this time sampling.

In addition, for each source, the UVES archive was queried to make use of all available and public data on the SACY targets.  
All data were reduced using the UVES pipeline recipe {\it uves\_obs\_redchain} with the command-line driven utility {\it esorex} (bias corrected, dark current corrected, flat-fielded, wavelength-calibrated and extracted). This outputs three individual spectra, (in the case of the standard set-up) one spectrum for each CCD chip of the instrument: blue, red upper and red lower (BLUE 3250-4500\,$\AA$, REDU 4800-5800\,$\AA$, and REDL 5800-6800\,$\AA$, respectively).  \\

In addition, the VizieR catalogue service was used to look for any existing RV values. A comparison between the archival data and the SACY values is shown in Figure~\ref{fig:lit_comparison}.  The individual values from UVES, FEROS and archival sources is shown in Table~\ref{table:av_values_all_sources} with their respective uncertainties; the average uncertainty from the literature is 0.9\,km\,s$^{-1}$.





\section{Characterising the sample}
\label{section:characterisation}

\subsection{Revising membership with new RV data}
\label{section:new_membership}


Accurately determined RV values are a key ingredient to the convergence method \citep{Torres2006} used to identify the SACY association members. The convergence method uses the gamma (systemic) velocity; in the case of single stars, this is equal to the RV.  For multiple systems without orbital solutions, there are two cases, with multi-component spectra (SB2/SB3) and without (SB1). For SB2/SB3 systems it is derived from weighting the component RV values, according to their relative flux.  For SB1 systems the mean RV value is used with its respective variability induced by further components (certainty of membership for these systems is very difficult without an orbital solution).  For multiple systems with orbital solutions it is a known value.  Therefore the more accurate the RV values and, in the case of multiple systems, the greater phase coverage observed, the more stringent the conditions for membership become.

The large number of RV values calculated in this work was used to revise and refine the member list \citep{dasilva2009}. With this new data any targets that no longer had a high probability of membership were removed from the overall analysis and statistics of the associations (initial members and revised members are shown in Table~\ref{tab:member_list}). However, as we have conducted useful analysis on these objects, details can be found in Appendix~\ref{sec:Individual Sources}.

\subsection{Mass distribution}
\label{section:Mass}

The mass distribution was constructed using effective temperatures from \cite{dasilva2009}.  First, available effective temperature were used directly with age estimates, as given in \cite{Torres2006}, to obtain a model-based mass using isochrones \citep{Baraffe1998, Baraffe2003, Siess2000}.  The different isochrones were used depending on the temperature range.  The median effective temperature for each spectral subclass within each association was used to estimate a value for those targets with a spectral type, as
calculated by the method described in \cite{Torres2006}, but without an effective temperature value.  Linear interpolation was used for those spectral subclasses without any corresponding temperature values from \cite{dasilva2009}.
In the case of Carina (CAR), effective temperature values for targets of spectral type K3 and later were taken from median values of Tucana-Horologium (THA), which are approximately the same age, and having a much larger population in this spectral type range. Masses were then derived using effective temperatures and isochrones to produce a mass distribution, a shown in Figure~\ref{fig:mass_dist}. 

\begin{figure}[h]
\begin{center}
\includegraphics[width=0.49\textwidth]{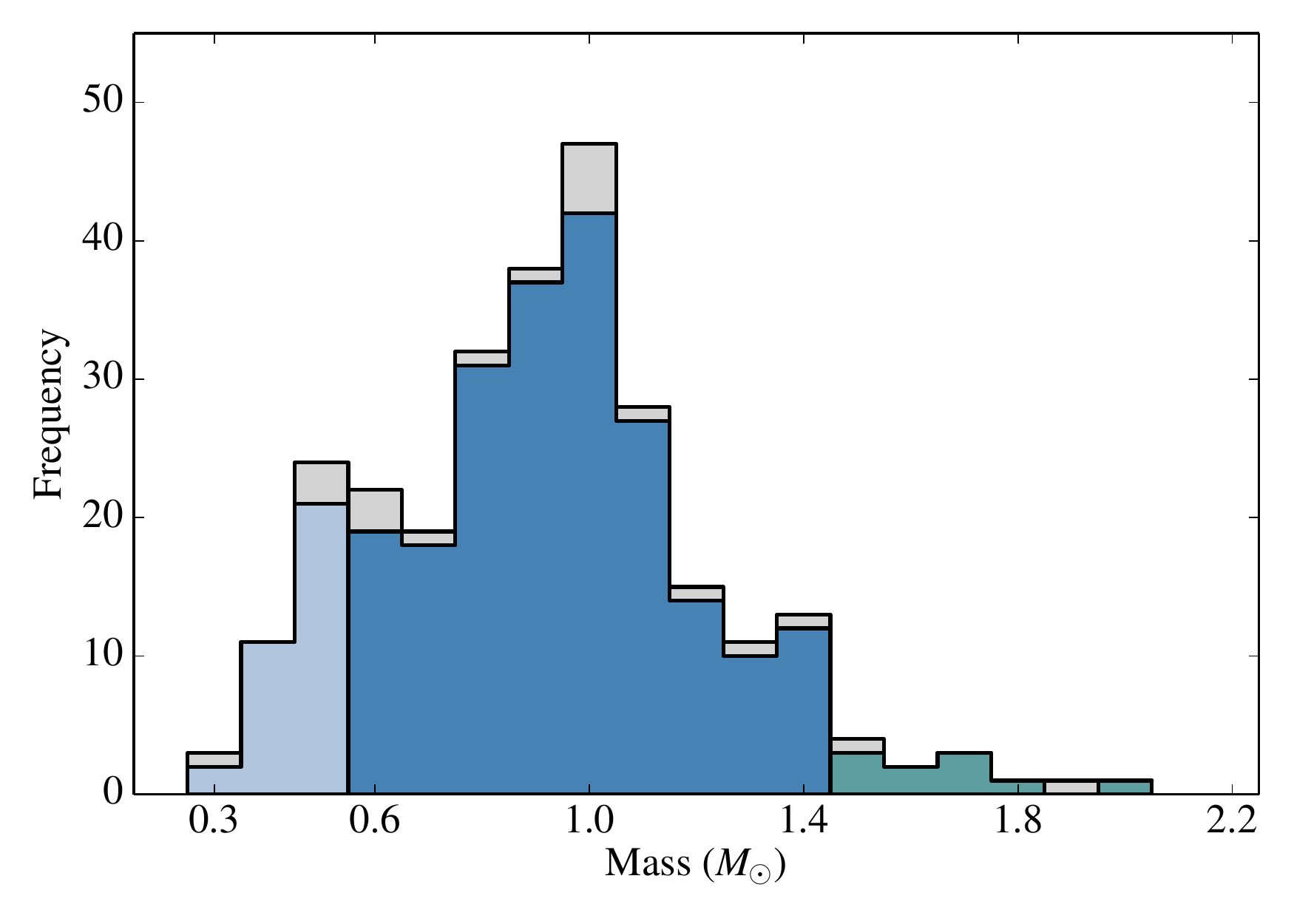}
\end{center}
\vspace{-0.5cm}
\caption{Mass distribution of all targets used in this work calculated by using available effective temperatures from \cite{dasilva2009} and derived masses using \cite{Baraffe2003} models $< 0.6\,M_{\odot}$ -- light blue,  \cite{Baraffe1998} models between masses 0.6-1.4\,M$_{\odot}$ -- dark blue, and \cite{Siess2000} models $>1.4\,M_{\odot}$ -- green. The grey represents multiple system primary masses.}
\label{fig:mass_dist}
\end{figure}

To validate this method we derived masses from a calculated bolometric luminosity value using the VOSA tool \citep{Bayo2008} by assuming a solar metallicity \citep{Almeida2009} and $log(g)$ between 3.5 and 5. This value is calculated from a fit of the Spectral Energy Distribution (SED).  Some
78 sources had poor SED fits due to a lack of available photometry and therefore could not be used any further.  The calibration of the bolometric luminosity is also based on the distance, which is the largest source of uncertainty.  This uncertainty combined with the photometric error count yields a further 65 sources (from $\approx$250) with an error in bolometric luminosity $>$20\% of the calculated value.  For the remaining sources, we derived the masses with the bolometric luminosity value using the isochrones.  The standard deviation in masses derived from the two different methods was 0.07\,$M_{\odot}$.  This is a very small and acceptable difference in our model-based masses.  

The mass distribution calculated from the effective temperatures was used in this work, as opposed to that calculated from the bolometric luminosity, as it represents our sample more fully.

\begin{figure*}[tbp]
\begin{center}
\includegraphics[width=0.98\textwidth]{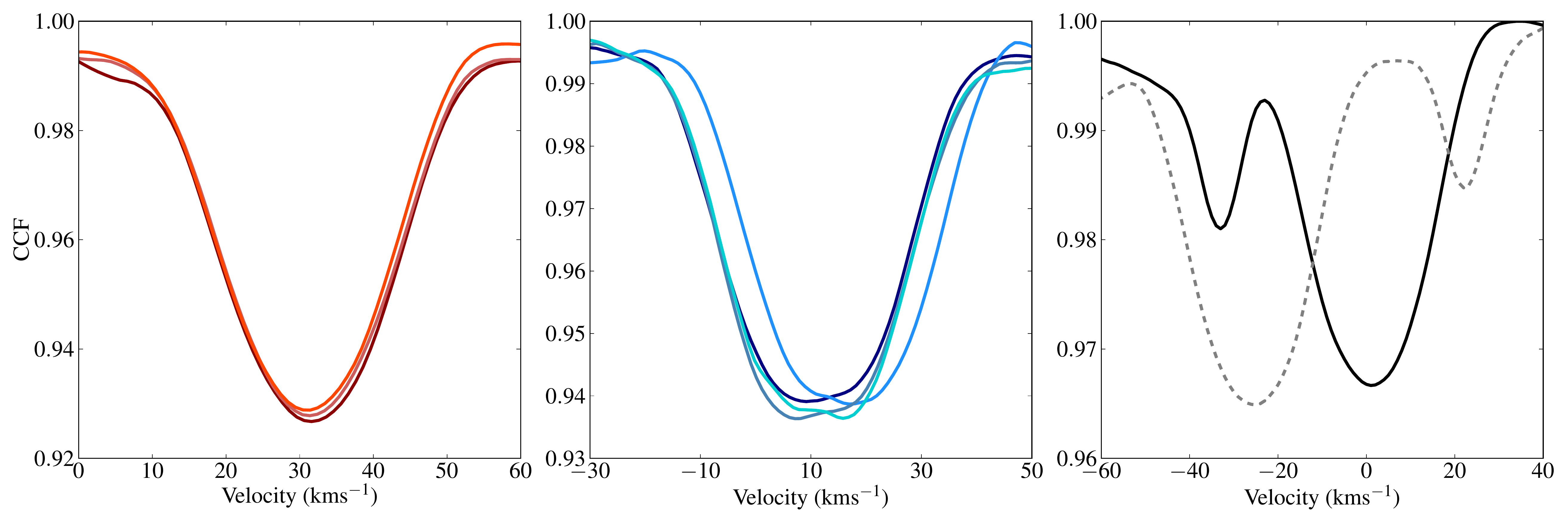}
\end{center}
\vspace{-0.5cm}
\caption{Resulting CCF profiles from three different sources. {\it Left Panel:} HD 45270 with no significant variation in RV and standard deviation $<$ 0.05\,km\,s$^{-1}$ (data epochs: 2007-05-08, 2008-12-05, 2010-09-26).  {\it Central Panel:} HD 104467 a SB1 candidate with significant variation in RV and std dev. $>2.70$\,km\,s$^{-1}$ (data epochs: 2009-03-16, 2010-05-26, 2012-02-24, 2012-03-07).  {\it Right Panel:} HD 155177 a SB2 system, as shown by two clear peaks and temporal movement (data epochs: 2007-04-19, 2012-05-09).}
\label{fig:ccf_profiles}
\end{figure*}





\section{Identifying multiple systems}
\label{sec:Identifying Multiple Systems}

To determine the RV values, we computed cross-correlation functions (CCFs) for all reduced spectra with a signal-to-noise ratio greater than 10. 
To compute the CCF, the observed spectrum is convolved with a CORAVEL-type numerical mask, as described in \cite{Queloz1995}.  The shape of the CCF function is approximated by a Gaussian profile and, in cases of fast rotation, the Gray rotational profile \citep{Gray1976}.   The RV is the peak of this profile to which the barycentric correction is applied.




For the sake of homogeneity, all CCFs quoted in this paper have been calculated using a K0 mask, including those without a literature spectral type determination.
Double-lined spectroscopic systems (SB2s) --and in rare cases, triple-lined spectroscopic system (SB3s) -- are identified visually, displaying two distinct peaks in the outputted CCF. For systems with indications of multiplicity in the CCF profiles, such as two apparently merged components, more research and analysis was conducted, the details of which can be found in Appendix~\ref{sub_sec:questionable_sb2}.  Examples of the CCF output for these systems and that of a system with no indication of multiplicity are shown in Figure~\ref{fig:ccf_profiles}.

\begin{figure}[h]
\includegraphics[width=0.48\textwidth]{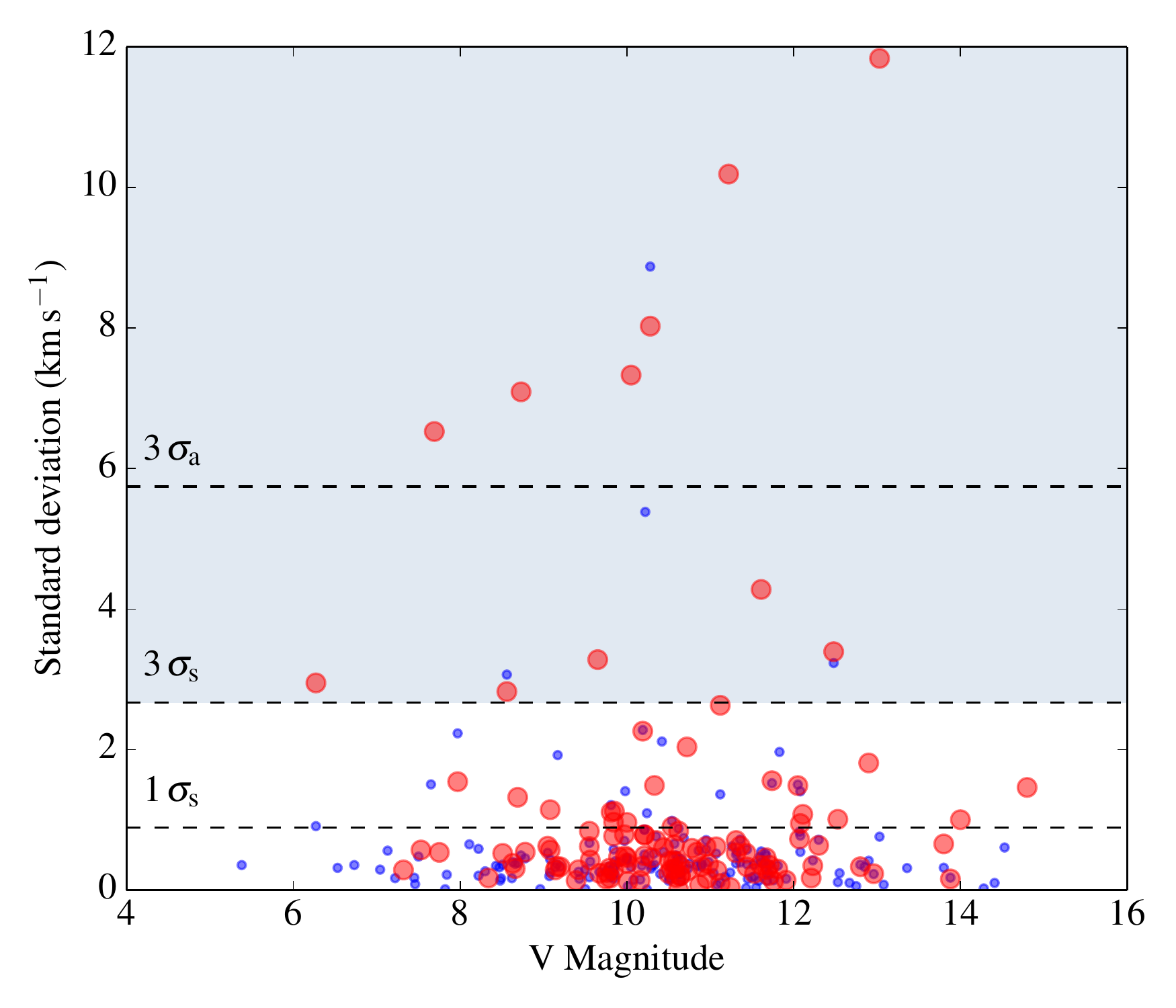}
\vspace{-0.2cm}
\caption{The standard deviation in RV values for SACY sources, excluding known SB2 systems, as a function of V magnitude.  UVES and FEROS values -- blue dots and UVES, FEROS and archival data -- red circles.The variables, 1$\sigma_\mathrm{s}$ and 3$\sigma_\mathrm{s}$, represent one standard deviation and three standard deviations respectively, for both UVES and FEROS data. The shaded area contains SB1 candidate multiple systems. The variable 3$\sigma_\mathrm{a}$, represents three standard deviations including archival data.}
\label{fig:figure9}
\end{figure}

Single-lined spectroscopic systems (SB1s) are identified through significant RV variations between different data epochs; the results are shown in Figure~\ref{fig:figure9}.  The shaded area contains all the SB1 candidates; these candidates have significant RV variations; above 3$\sigma_\mathrm{s}$ (2.70\,km\,s$^{-1}$), or three standard deviations for the entire dataset.  The value of three standard deviations, 3$\sigma_a$, including archival data is also shown for comparison.  All candidates above 3$\sigma_\mathrm{s}$ were investigated further, details can be found in Appendix~\ref{subsec:archival_rv_variations}.

\subsection{Sources of uncertainty}

The uncertainty in the RV values is calculated from equation (9) of \cite{Baranne1996}:
\begin{equation}
\sigma_{\mathrm{meas.}}={C(T_{\mathrm{eff}})\over D \times S/N}{(1+0.2\omega)\over3}~\mathrm{km\,s}^{-1},\end{equation}

where $C(T_{\mathrm{eff}})$ is a constant that depends on both the spectral type of the star and the mask used, which is typically 0.04, $\omega$ is the (noiseless) FWHM (km\,s$^{-1}$) of the CCF, $D$ is its (noiseless) relative depth, and $S/N$ is the mean signal-to-noise ratio.

This produces a value, which is the internal uncertainty for the measurement of the RV using this CCF method.  Due to the very high signal-to-noise ratio in our data (average $\sim$ 100), the contribution of this internal uncertainty is almost negligible. Although the value of $C(T_{\mathrm{eff}})$ is a function of the $T_{\mathrm{eff}}$ and the mask used to compute the variation in the temperature range 4000-6000\,K, it only changes the value of $C(T_{\mathrm{eff}})$ by $\approx$0.01.  When this variation is propagated across the $T_{\mathrm{eff}}$, the output range is not significantly affected.
Values are shown in Table~\ref{tab:master_individ_rv}, these typical values of the uncertainty are $\sim$0.01\,km\,s$^{-1}$.  However, the stars in our study are often variable, and this can induce deformities into the CCF \citep{Lagrange2013} in which this calculation of uncertainty does not account for; it assumes a symmetric CCF profile, and thus, the uncertainty is underestimated in the majority of cases.

A more empirical approach to gauging the level of uncertainty in the measurements is to use the standard deviation of the RV values, as shown in Figure~\ref{fig:figure9}.  This is an {\it a posteriori} approach, assuming a very small percentage of our candidates are in multiple systems, and that this overall distribution is mainly representing single systems.  Previous research in SFRs concerning short period multiplicity of T-Tauri stars has shown that single systems are by far the most abundant 90+\% \citep{Melo2003} and have used such distributions to estimate empirical uncertainties. The one sigma level in RV variation for all targets with more than one data epoch (excluding known SB2 systems) is 0.89\,km\,s$^{-1}$, as seen in Figure~\ref{fig:figure9}.



\subsection{Accounting for biases}
\label{subsec: bias}

A huge advantage of the SACY dataset is the way in which the sample was compiled from optical counterparts to X-ray sources from the ROSAT all-sky survey.  X-ray emission can originate in protostars $\sim10^4-10^5$ yr \citep{Koyama1996} to post T-Tauri stars $\sim$10 Myr \citep{Walter1988}.  In terms of mass this is from sub-stellar \citep{Neuhauser1996} to intermediate-mass Herbig Ae/Be stars \citep{Zinnecker1994}.  On this basis, our sample should be relatively free of significant bias for ages and  primary masses ($\sim$5-70 Myr and 0.5-3.0 $M_{\odot}$).  However, the sample is not bias-free regarding the characteristics of the secondary components and orbital parameters.





To estimate the completeness of this work, given the sparse and irregular time sampling, in terms of our sensitivity in observing binary systems, probability-detection density maps were created using a set of synthetic binary systems, as described in Section 6.1 of \cite{Duquennoy1991} and outlined further below.  The primary masses used in these simulations are the mean values of mass bins containing equal numbers of targets {(0.6, 0.9 and 1.2 $M_{\odot}$: 85 targets) taken from the overall mass distribution of the sample, which includes all targets with cross-correlation results not previously identified as SB2 systems.

For each primary mass, a density map as a function of the secondary mass ($M_2$) and the period ($P$) of the binary system was constructed. A set of binary systems are generated in this space ($M_2/M_1$: 0.1M$_{\odot}$ to $M_1$, $P$: 1 to 1000 day) following certain distributions.  The parameters phase at time t=0 ($\phi_{t=0}$), longitude of the ascending mode ($\omega$) and the inclination ($i$), are randomly drawn from uniform distributions within their respective numerical ranges. The eccentricity ($e$) values depend on the period of the binary system as follows:
\begin{itemize}
\item $P< 8$ day, $e\equiv0$
\item $P\le8<1000$ day, $e$ is randomly drawn from a normal distribution centred at 0.33 with a standard deviation of 0.03 \citep{Mayor1984}.
\end{itemize}

The semi-amplitude ($K$) and thus the RV were then calculated through equations (14) and (17) from \cite{Halbwachs2001} using our physical set of data epochs to input the time ($t$) into the equations. 
For each object, a set of RVs is produced, and from this, the standard deviation from the mean is calculated. If this calculated value is above the standard deviation of our physical, observed data the system is said to be {\it observed}; this process is then repeated $N$ times for each ($M_2, P$) point.  


\begin{figure*}[tbp]
\begin{center}
\includegraphics[width=1.0\textwidth]{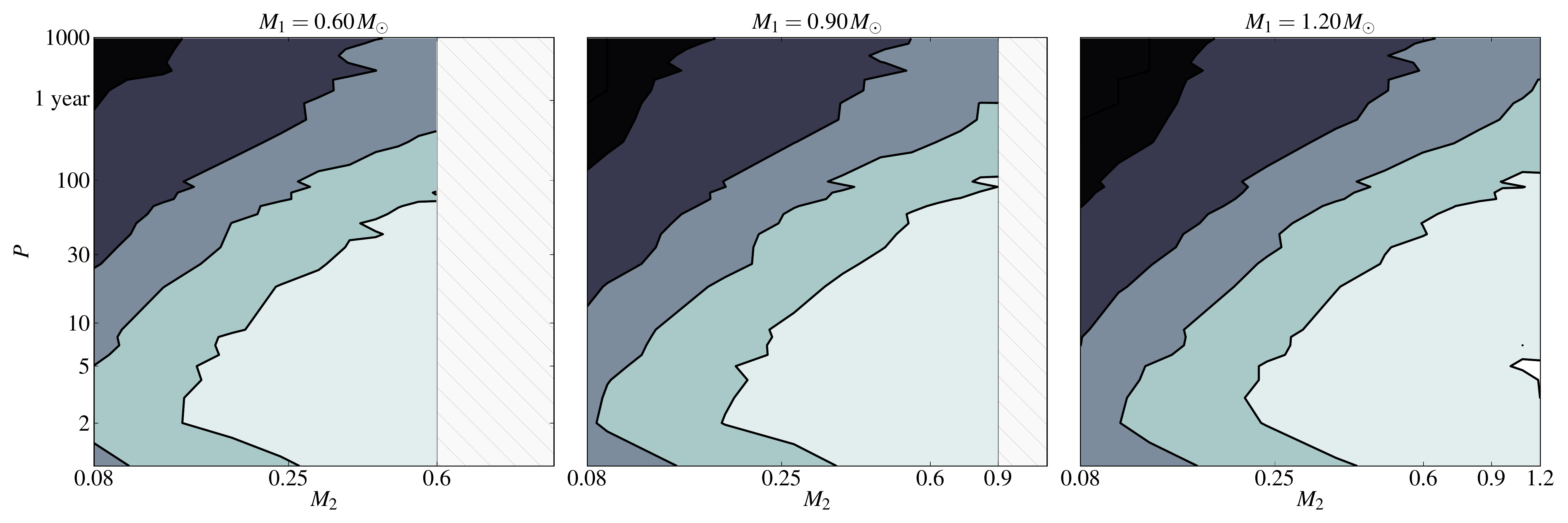}
\end{center}
\caption{Probability density maps in ($M_2$, $P$) space for three different primary mass stars, 0.60, 0.90, and 1.20\,$M_{\odot}$, periods, $P$, range from 1-1000\,day, and secondary masses, $M_2$, from $0.1\,M_1 - M_1$.  Probabilities: $<$ 5\% -- black, 5-39\%, 40-59\%, 60-74\%, 75-90\% and $>$ 90\% -- white.}
\label{fig:figure6}
\end{figure*}

For this range Figure~\ref{fig:figure6} shows that the effect of the primary mass of the binary system has little effect on the sensitivity to observe the secondary.
  For all three primary masses considering a secondary component of 0.1\,$M_{\odot}$, orbital periods of 30 day and 100 day have a $>$75\% and $>$60\% probability, respectively, of being detected within our dataset.

\section{Results and discussion}
\label{section:Results}


The individual targets classified as multiple systems are shown in Table~\ref{table:all_sbs}, which are identified either as SB1, SB2 or SB3.  We then calculate the multiplicity fraction ($F_\mathrm{m}$):

\begin{equation}
F_\mathrm{m}={B+T+Q\over S+B+T+Q},
\end{equation}

where $S$, $B$, $T$, and $Q$ are the number of single, binary, triple, and quadruple systems, respectively.\\

We use the multiplicity fraction, as opposed to the Companion Star Fraction (CSF), which is the number of companions per star, because the multiplicity fraction is more robust against unobserved higher order components \citep{Hubber2005}, whether a multiple system is a binary or a triple system the overall value of $F_\mathrm{m}$ would not be affected.

{\tiny
\begin{table}
\caption{Targets classified as multiple systems.}
\begin{tabular}{p{3.2cm} p{0.6cm} p{0.5cm} p{0.7cm} p{0.7cm} p{0.7cm}}
\hline\hline\\
  \multicolumn{1}{l}{ID} &
  \multicolumn{1}{l}{Mass} &
  \multicolumn{1}{l}{SpT} &
  \multicolumn{1}{l}{Ass.} &
  \multicolumn{1}{l}{Age} &
  \multicolumn{1}{l}{Flag} \\
  \multicolumn{1}{l}{} &
  \multicolumn{1}{l}{($M_{\odot}$)} &
  \multicolumn{1}{l}{} &
  \multicolumn{1}{l}{} &
  \multicolumn{1}{l}{(Myr)} &
  \multicolumn{1}{l}{} \\[1ex]
\hline\\
  GSC 09420-00948 & 0.6 & M0 & ECH & 6 & SB1\tablefootmark{a} \\
  HD 104467     & 1.9 & G3 & ECH  & 6 & SB1\tablefootmark{a} \\
  V1005 Ori     & 0.7 & M0 & BPC & 10 & SB1\tablefootmark{b} \\
  HD 59169      & 1.05 & G7 & ABD & 70 & SB1\tablefootmark{a} \\
  PX Vir               & 0.9  & K1 & ABD & 70 & SB1\tablefootmark{b} \\
  CD-2711535   & 1.0  & K5 & BPC & 10 &  SB1\tablefootmark{a,b} \\
  AK Pic              & 1.2 &  G2 &  ABD & 70 & SB1\tablefootmark{b} \\
  GSC 08077-01788 & 0.6 &  M0 &  COL & 30 & SB1\tablefootmark{a,b} \\
1RXS J195602.8-320720  & 0.3 & M4 & BPC & 10 & SB2\tablefootmark{a} \\
  V4046-Sgr    & 1.1 & K6 & BPC & 10 & SB2 \\
  HD 155177     & 1.5 & F5 & OCT & 20 & SB2\tablefootmark{a} \\
  BD-20951     & 1.0 & K1 & CAR & 30 & SB2 \\
  HD 309751     & 1.05 & G5 & CAR & 30 & SB2 \\
  TWA 20\tablefootmark{*}    & 0.4 & M3 & TWA & 8 & SB2\\
  CD-423328  & 1.0 & K1 & CAR & 30 & SB2\tablefootmark{b,c}\\
  HD 217379     & 0.6 & K7 & ABD & 70 & SB3 \\
  HD 33999      & 1.3 & F8 & ABD & 70 & SB3 \\[1ex]
\hline
\end{tabular}
\label{table:all_sbs}
\tablefoot{\tablefoottext{*}{No longer classified as a member of TWA with new RV data.}\tablefoottext{a}{System has not been previously identified in related SACY work or literature.}\tablefoottext{b}{Targets classified based on variation from literature RV values.}  \tablefoottext{c}{Details of the technique to classify this target are described in section~\ref{sub_sec:questionable_sb2}.}}

\end{table}
}









\subsubsection{Multiplicity fraction as a function of age}

The SACY sample has age determinations calculated in \cite{Torres2008}.  However, there are significant uncertainties within these determinations (as shown in Table~\ref{table:summary_associations}), which are derived using a convergence method represented by third degree polynomials of $(V-I)_C$.  There is an abundance of papers on the age determination of young associations, including the well-known $\beta$-Pic: $12^{+8}_{-4}$\,Myr \citep{Zuckerman2001}, $20\pm$10\,Myr \citep{Barrado1999}, TW-Hydrae: $10^{+10}_{-7}$\,Myr \citep{Barrado2006}.  This is often a controversial subject; however, the relative ages between the associations are less affected and, therefore, are more robust quantities.  In other words, it is known that TW-Hydrae is younger than Tucana-Horologium, irrespective of their absolute ages, as seen in \cite{dasilva2009}, Figure 3.  

If there was a significant relationship between multiplicity and age (between ages $\sim$5-70\,Myr) we could observe it. However, as shown in Figure~\ref{fig:figure10}, we see no such trend, merely scatter across the age range.  These results also agree with data from Young clusters, Tau-Aur and Cha I with ages 1\,Myr and 2\,Myr, respectively \citep{Luhman2004}.  In addition to this when considering results from the field within the orbital period range (1-200\,day), the fraction, $0.073^{+0.014}_{-0.012}$, is indistinguishable \citep{Duquennoy1991, Nguyen2012}.

These results can be explained by the minimal N-body dynamical processing for systems with small physical separations ($<$50\,A.U.).  Such systems are essentially never destroyed \citep{Parker2009}, and if we assume similar numbers of multiple systems are created from population to population initially, we would not expect to see a significant relationship with time.


\vspace{-0.2cm}
\begin{figure}[h]
\begin{center}
\includegraphics[width=0.49\textwidth]{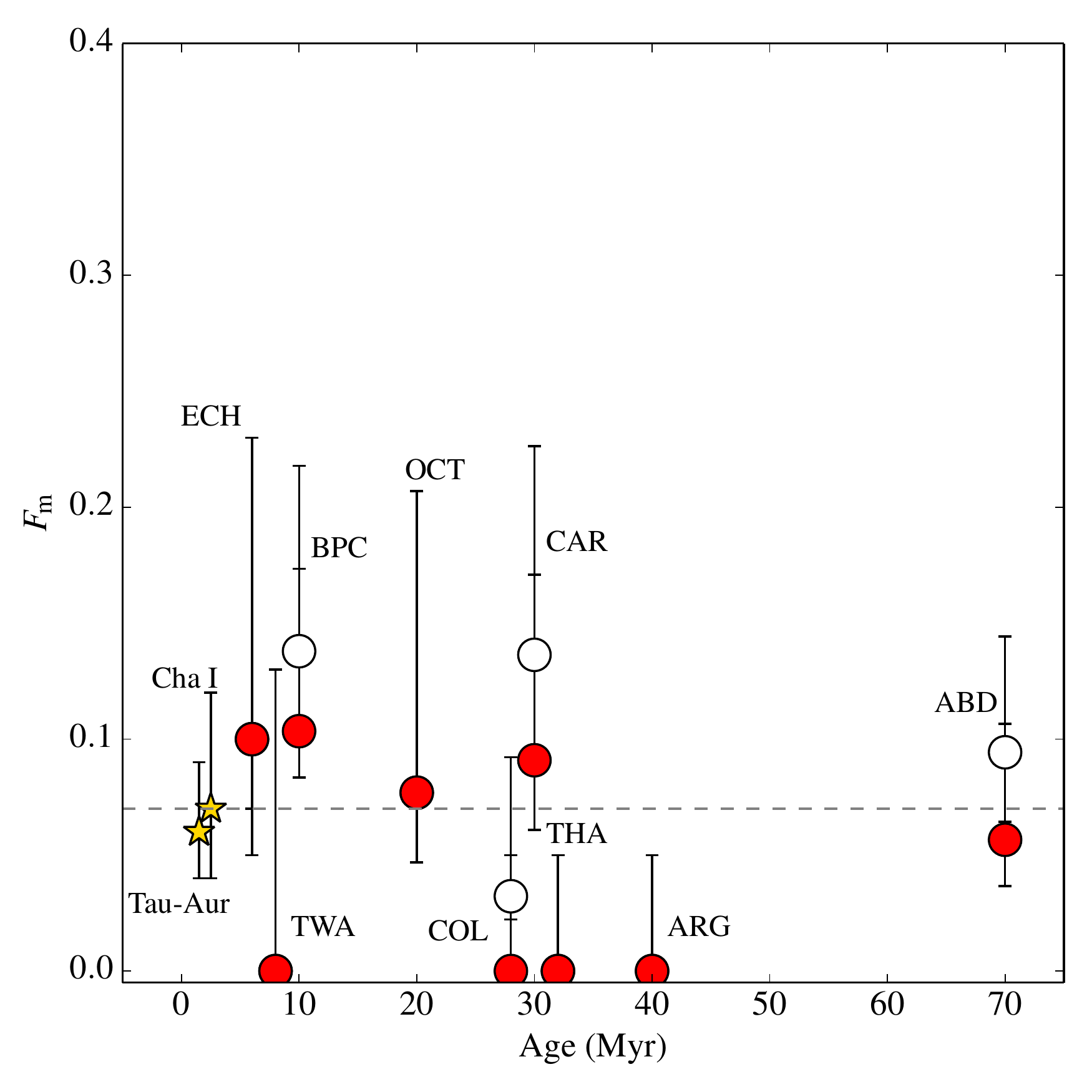}
\end{center}
\vspace{-0.5cm}
\caption{The fraction of spectroscopic multiple systems from the overall observed sample as a function of age for each of the SACY associations, which are individually labelled.  UVES and FEROS data -- red, UVES, FEROS and archival data -- white, and data from \cite{Nguyen2012} -- yellow.  Uncertainties were derived from low-number statistics using Equation (A3) from \cite{Burgasser2003}. A line representing 7\% multiplicity is shown.}
\label{fig:figure10}
\end{figure}

\subsubsection{Multiplicity fraction as a function of mass ($< 3 M_{\odot}$)}
\label{subsubsection: MF with Mass}


To compare our results in terms of primary mass to previous work, we must define a consistent period and secondary mass range and apply the same correctional techniques.  \cite{Nguyen2012} studied the range $P_{\mathrm{min}}$=1, $P_{\mathrm{max}}$=200\,day and $M_{2, \mathrm{min}}$=0.08, $M_{2, \mathrm{max}}$=0.6\,$M_{\odot}$.  The properties of the synthetic binaries used in their simulations 
are taken from a typical T-Tauri star in their sample; however we have calculated detection probabilities using three primary masses (0.6, 0.9 and 1.2\,$M_{\odot}$) in this work.  

We can estimate the probability that the multiple system will be in this $P$ and $M_2$ range from each primary mass $M_1$ using Bayes' theorem \citep{Bayes1763}:

\begin{equation}
\mathcal{P}_{M_1}(m , p)={\sum_{m=M_2, \mathrm{min}}^{M_2, \mathrm{max}} \sum_{p=P_{\mathrm{min}}}^{P_{\mathrm{max}}} \mathcal{M}(m , p)\mathcal{D}(m,p) \over \sum_m\sum_p\mathcal{M}(m , p)\mathcal{D}(m,p)},
\end{equation}

where $\mathcal{M}(m , p)$ is the prior distribution.  In this case the most ignorant prior distribution is used, $\mathcal{M}(m , p)=1$, and $\mathcal{D}(m,p)$ is the likelihood function from our probability of detection simulations. 

We compute the probability for each of our primary masses and use the mean detectability probability to compute a corrected fraction, as below:

\begin{equation}
F_{\mathrm{correc}}={F_{\mathrm{obs}}\times \mathcal{P}_{M_1}(m , p) \over {1 \over N} \sum_{m=M_2, \mathrm{min}}^{M_2, \mathrm{max}}\sum_{p=P_{\mathrm{min}}}^{P_{\mathrm{max}}}\mathcal{D}(m,p)},
\end{equation}

where $F_{\mathrm{correc}}$ is the corrected fraction, $F_{\mathrm{obs}}$ is the observed fraction, and $N$ is the number of separate detection probabilities from the simulations.  

The results of these corrected fractions are shown in Figure~\ref{fig:figure34}, along with data from \cite{Nguyen2012}. Our mass bins contain equal numbers of targets (85) to eliminate any bias from the frequency of objects in a certain mass range. 
The results are fully compatible, and the spread in the value of $F_{\mathrm{correc}}$ including the uncertainties is only 7\% across the mass range ($\approx$0.2--2.0~$M_{\odot}$).  This result also supports the universality of multiplicity, as the data from the two different environments (loose associations and denser SFRs) is indistinguishable.

As well as comparing the results from SACY to those of younger SFRs, one can look at the form of the corrected fraction with primary mass for both environments.   Figure~\ref{fig:figure34} shows there is no significant increase or decrease in the corrected fraction as a function of the primary mass.  In other words, from our results, it is just as likely to observe a multiple system with a primary mass of $\approx$0.2$M_{\odot}$ than one with a primary mass of $\approx$2.0$M_{\odot}$.  As shown and discussed in Section~\ref{subsubsection: MF with Mass 2} when higher masses are included the form changes quite dramatically.

\vspace{-0.2cm}
\begin{figure}[h]
\begin{center}
\includegraphics[width=0.49\textwidth]{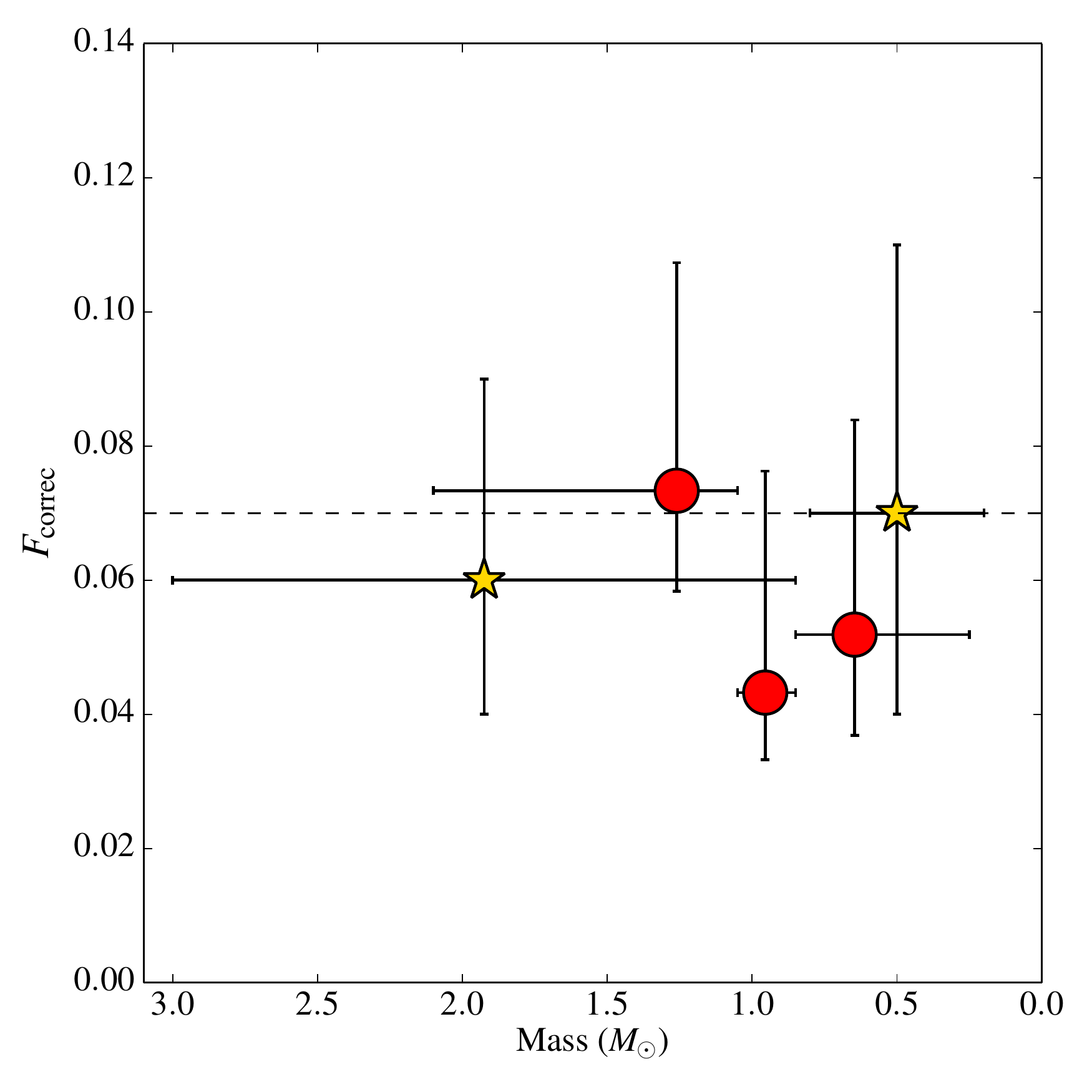}
\end{center}
\vspace{-0.5cm}
\caption{Multiplicity fraction as a function of primary mass; mass bins are represented with x-axis error bars with the average value for the SACY targets being the red markers.  Fractions have been corrected using the techniques outlined in Section~\ref{subsubsection: MF with Mass}.  The same data \citep{Nguyen2012} shown in Figure 6 has been transformed from spectral type to mass by assuming an age of 1 Myr and 2 Myr for Tau-Aur and Cha I respectively, using $T_{\mathrm{eff}}$ from \cite{Luhman2004} and \cite{Sestito2008} with isochrones explained in Figure ~\ref{fig:mass_dist}.}
\label{fig:figure34}
\end{figure}




%

\subsubsection{Multiplicity fraction across a wide mass range}
\label{subsubsection: MF with Mass 2}

There are two phases of interaction that can shape the multiplicity fraction we observe in populations of stars: (I.) Interactions within the pre-stellar cores ($\leq10^{5}$\,yr) through accretion, disk interaction and the dynamics of multiple systems \citep{Bate2002} and (II.) through cluster dynamics, where the interactions between stars that have originated in different pre-stellar cores ($>10^5$\,yr), which can be modelled using N-body simulations \citep{Kroupa1995}.


Populations that are $\sim$Myr have undergone significant evolution of their parameters already (e.g., mass ratio, separation distribution, and core multiplicity fraction).  However, the close systems observed in this study ($<10$\,A.U.) should not have undergone any significant processing through the second phase described above.  Therefore, even across a wide range of ages and environments, we should be probing the multiplicity fraction that is a direct result from the evolution within the separate pre-stellar cores.  


Figure~\ref{fig:sb_spec} shows a compilation of results from this study (in red and white) and from the literature (in blue and gold) for spectroscopic multiplicity fractions from a range of sources.  The apparent exponential form of this function should be considered with caution in this instance, as this compilation lacks thorough refinement, such as a defined consistent orbital period range. It is a qualitative example of the trend of increasing multiplicity fraction with spectral type, and therefore mass, for tightly bound multiple systems. This trend seems to be preserved when considering much wider systems in the field, as seen in Figure 12 of \cite{Raghavan2010}.  

One potential explanation for this observed relationship in the field relates to the binding energy of the system; however, we would not see a significant relationship for tight binaries on this basis,
(\cite{Parker2009}, Section 5.2, interaction time-scale calculations: \cite{Nguyen2012}).  As shown in Figure~\ref{fig:sb_spec}, there is an increase with mass for these close systems, this is not easily explained with our current understanding of star formation. 


The observed physical separation of close systems ($<$10\,A.U.) is the result of the evolution of the multiple system's properties. The size of the first Larson core, a pressure-supported fragment of $\approx$ 5\,A.U. \citep{Larson1969}, does not permit a secondary component to orbit at such a physical separation initially.  \cite{Bate2002} shows how these close systems can be formed from wider systems through the processes associated with phase I with dynamical interactions being the dominant mechanism. 
%
%
Kozai cycles with tidal friction (KCTF) \citep{Eggleton2006} is an example of such a dynamical process.  This evolution depends on the periods of the primary and tertiary components $P_3(P_3/P_1)$. Given $P_1$= 1\,yr and $P_3$ = 1000\,yr, the time-scale for one cycle is $\sim10^6$\,yr; $>10^2$ cycles are needed to complete the evolution \citep{Tokovinin2006}.  Therefore, one could argue that this evolution is negligible at the PMS stage.  However, the periods within the system could be initially much more similar and thus the ratio would be smaller, decreasing the time-scale.  On this basis, this mechanism could have an effect even at the PMS stage, producing SB systems.

\cite{Bate2002} conclude that the processes described in phase I favour the production of higher mass multiple systems, which would explain the trend shown in Figure~\ref{fig:sb_spec}.  If these processes are responsible for the systems we observe, there would be other related properties, such as a lack of extreme mass-ratios of the systems and a tendency for wider companions.  Both of these properties will be investigated in further SACY work: Mass ratios by following up identified spectroscopic candidates to characterize their orbital parameters and wider companions by analysis and inclusion of already existing adaptive optics (AO)-assisted direct imaging data.

\begin{figure}[h]
\begin{center}
\includegraphics[width=0.49\textwidth]{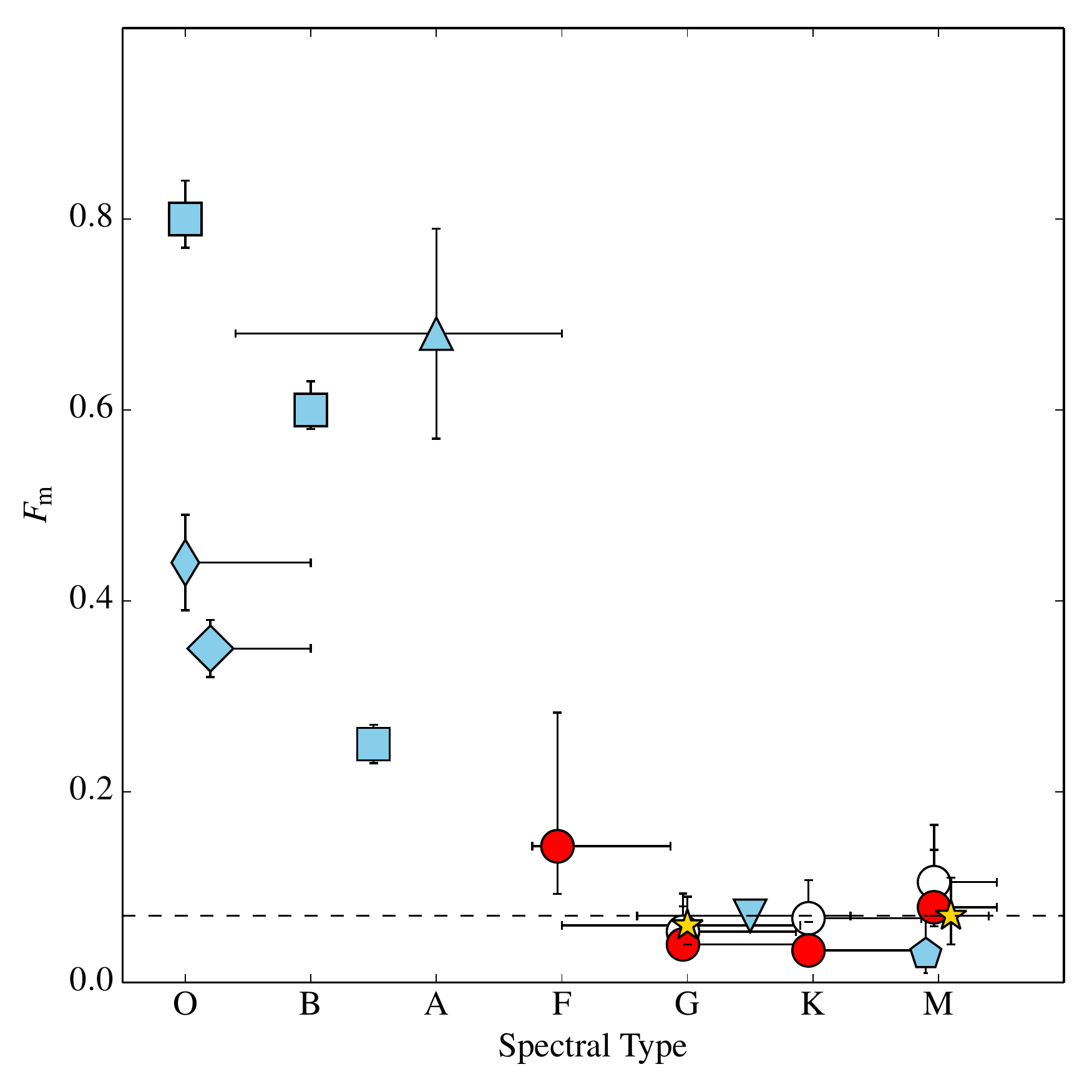}
\vspace{-0.5cm}
\caption{The spectral type versus the multiplicity fraction for all available SACY data and studies from the literature; the horizontal error bars represent the width of the spectral class bin (symbols in red, white and yellow as described in previous figures). \cite{Chini2012} -- filled square, \cite{Baines2006} --  filled triangle, \cite{Sana2011} -- filled thin diamond, \cite{Sana2013} -- filled diamond, \cite{Raghavan2010} -- inverted filled triangle, and \cite{Fischer1992} -- filled pentagon.}
\label{fig:sb_spec}
\end{center}
\end{figure}


One must also consider the possibility that star formation is not a single universal process.  \cite{King2012b} have found significant statistical differences\footnote{see~\cite{Marks2014} for an alternative interpretation} in the multiplicity fraction for binary systems in the 19-100\,A.U. range (dynamically unprocessed) between the field and five clusters (Taurus, Upper Scorpius, Corona Australis, Chamaeleon I, and Ophiuchus). They conclude that the star formation process is non-universal and different environments produce different numbers of binaries due to this. 

In our data presented in Figure~\ref{fig:figure10}, we find three associations that show no spectroscopic systems from our current dataset: Argus: 0 / 33 ($0^{+7}\%$), Tucana-Horologium: 0 / 33 ($0^{+5}\%$), and TW-Hydrae: 0 / 11 ($0^{+13}\%$). With our current dataset this means that Argus, Tuc-Hor, and TW-Hydrae have upper limits on the $F_{\mathrm{m}}$ value of 0.07, 0.05, and 0.13, respectively.  These results could also be explained by the non-universality of formation; however, these are still very low numbers and are not statistically significant.

There is an alternative theory to explain this apparent $spread$ in binary fractions (and properties): turbulent fragmentation. \cite{Jumper2013} describe how binary properties can be determined very early on by turbulence within the pre-stellar cores; this extremely stochastic process intrinsically leads to a range of binary properties and as a product to the formation of the Brown Dwarf (BD) desert.  Therefore, one should be careful when considering multiplicity fractions as a direct key to star formation, as we still need to clarify the picture further at the early stages of formation.  Work is ongoing in this area.




\section{Conclusions and further work}
\label{sec:Conclusions and Further Work}

\begin{enumerate}
\item Using our new RV determinations, we have revised the membership of the SACY associations.
\item We have identified seven new multiple systems (SB1s: 5, SB2s: 2) within the SACY sample, as shown in  Table~\ref{table:all_sbs}.

\item Within the nine SACY associations studied in this work, we have found that there is no significant age dependence on the observed fraction of spectroscopic multiple systems ($F_{\mathrm{m}}$).
\item The corrected fraction of spectroscopic multiple systems ($F_{\mathrm{correc}}$) within the mass range $\approx$0.2--2.0$~M_{\odot}$ is compatible with a flat distribution.
\item Our results are consistent, considering fractions as a function of mass and age, with nearby SFRs and the field, as expected from the picture of universal star formation.
\item The observed fraction of spectroscopic multiple systems sharply declines as a function of spectral type and appears to plateau for later-types.  To what spectral types this flat distribution extends still needs to be explored.

\end{enumerate}	




This study has analysed the fraction of close multiple systems through spectroscopic techniques.  In the future, these results will be complimented with AO-assisted direct imaging from high-angular resolution data.  Combining the analysis presented in this paper with that of much wider companions, we can get a wealth of extremely valuable information about the multiple systems in these young close loose associations and their similarities / differences to other environments in the context of star formation.  We can investigate certain multiple star formation scenarios by the detection / non-detection of wider tertiary components to the SB candidates proposed in this paper.  In addition to this, we will follow up our spectroscopic candidates identified in this study to better characterize their properties such as orbital period and mass ratio.\\


We have analysed $\approx$ 2700 spectra with an average S/N of 100 and succeeded in detecting new spectroscopic systems with our data alone and in combination with catalogue data.  From our literature search, we did not fail to detect any systems previously identified as multiple systems.  However, there is the possibility that we were not able to identify some multiple systems through our analysis. Our simulations, as seen in  Figure~\ref{fig:figure6}, are one method to try and account for possible biases in our dataset. However, it is important to note that the raw fractions, $F_\mathrm{m}$ as quoted in this paper, should be treated as lower limits, as 64/250 of our targets have only one RV value. \\



\begin{acknowledgements}

We would like to thank the anonymous referee for useful comments that improved this manuscript. 
Also to Isabelle Baraffe for vital feedback, to Matthew Bate and Gilles Chabrier for helpful and insightful discussions regarding star formation and multiplicity and to J\'er\^{o}me Bouvier and Estelle Moraux for their input and support of this work and future projects.
This publication has made use of VOSA, developed under the Spanish Virtual Observatory project supported from the Spanish MICINN through grant AyA2008-02156. Also the SIMBAD database and VizieR catalogue access tool, CDS, Strasbourg, France.
\end{acknowledgements}


\begin{appendix}
\section{Temperature scale}
\label{appendix:temp}

The way in which the mass distribution (Figure~\ref{fig:mass_dist}) of our sample was determined was based on effective temperature values in combination with model predictions.  When there was no effective temperature value from \cite{dasilva2009}, we used the spectral type of that target as an indication for its effective temperature and took the median value for all available targets with the same spectral type in that age bin.  If there were no other targets within that spectral type bin a value was constructed through linear interpolation, using the nearest earlier- and later-spectral type.  Figure~\ref{fig:t_eff_scale} shows the range of effective temperature and spectral type values.  

The spread in effective temperature in each spectral type bin, as shown by the grey shaded areas, is generally small across all three age bins, for example, for the later-types, which is smaller than the spread for younger objects reported in \cite{Bayo2011}, which are up to $\pm$ 150\,K.  The largest apparent spread is for the G3 spectral type bin at $<$20\,Myr; it is caused by two values: 4722\,K and 5759\,K.  The result is anomalous, and therefore, any necessary interpolation within the regime G2-G8 was conducted linearly using the two effective temperature values of 5988\,K and 5344\,K as boundaries, which is in accordance with the trend of the data.  

\begin{figure}[h]
\includegraphics[width=0.48\textwidth]{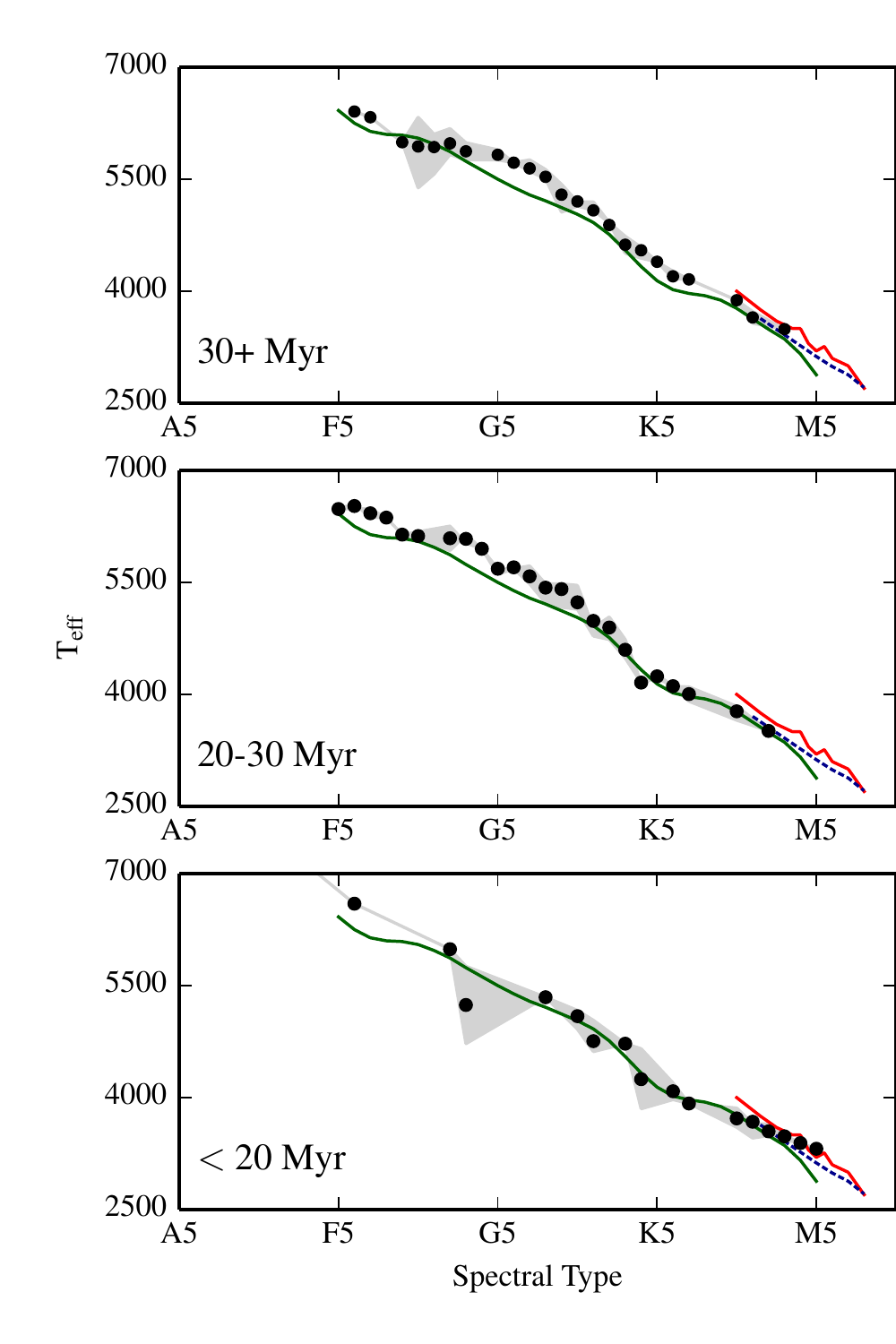}
\vspace{-0.5cm}
\caption{Spectral type versus the effective temperature for all stars used in this analysis.  The stars have been binned in three different age bins, as indicated in each panel. The grey shaded area represents the range of values within each spectral bin; the black marker is the mean value.  Also plotted are the results of \cite{Pecaut2013} -- green line, \cite{Luhman2003} -- dotted blue line, and \cite{Bayo2011} -- solid red line for comparison.}
\label{fig:t_eff_scale}
\end{figure}



\section{Notes on individual sources}
\label{sec:Individual Sources}

In Appendix~\ref{subsec:archival_rv_variations} we comment on the individual sources that show RV variation above $\sigma_\mathrm{s}$, 0.89\,km\,s$^{-1}$. Appendix~\ref{sec:Archival} shows the results from an archival search for any existing multiplicity flags. In Appendix~\ref{sec:def_sb2_systems} we comment on the data used to identify SB2 and SB3 systems in this work. Appendix~\ref{sub_sec:questionable_sb2} describes further analysis conducted in the cases of potential SB2/SB3 systems to remove false detections from deformities in the CCF profiles.

\subsection{Archival RV values}
\label{subsec:archival_rv_variations}
Values quoted below, calculated in this work, are average values from all available UVES and FEROS data; their uncertainties are one standard deviation in the following form: RV $\pm$ $\sigma $\,km\,s$^{-1}$. \\




\noindent{\bf HD 59169}:  There are only two available epochs for this object: one is from UVES data and the other is FEROS data producing a value of 34.3$\pm5.4$\,km\,s$^{-1}$.  However, this is a significant change in RV, and therefore, it remains a potential multiple system. \\


\noindent{\bf AU Mic}:  There are four data epochs in our own dataset, three from UVES, one from FEROS (-5.1,- 5.2, -4.9, and -4.0\,km\,s$^{-1}$) producing a value of -4.8$\pm$0.5\,km\,s$^{-1}$, which is not a significant variation. If one looks at the values from the literature (1.2$\pm$1.3, 15.6 and -4.5$\pm$1.3\,km\,s$^{-1}$.) there is one value that is significantly different from the others, 15.6.  However, after private communication with John E. Gizis, it is apparent that the 1$\sigma$ uncertainties are $\approx$17\,km\,s$^{-1}$, making this value compatible with all others.  Therefore, this target is not considered as a multiple system candidate. \\
 
\noindent{\bf V1005 Ori}:  This target has a value of 19.0$\pm$0.3\,km\,s$^{-1}$ from three UVES epochs, which is a variation well below 1$\sigma_\mathrm{s}$.  However, there are seven values from the literatures; which when evaluated together show clear RV variability.  Therefore, this target is considered a multiple system candidate. \\


\noindent{\bf GSC 09420-00948}: There are four UVES data epochs for this target, producing a value of 12.7$\pm$3.2\,km\,s$^{-1}$.  In combination with this, there are two values from \cite{Malaroda2006}: 5.0 and 17.1\,km\,s$^{-1}$.  However, these do not have any quoted uncertainties.  Because the variation in RV value is larger than $\sigma_\mathrm{s}$, even without the use of archival data, it remains a potential multiple system.  \\

\noindent{\bf HD 104467}: This target has a value of 13.3$\pm$3.1\,km\,s$^{-1}$ from both UVES and FEROS data.  There are also two values from VizieR: 12.3$\pm$1.4  \citep{Torres2006} and 15.4$\pm$1.0\,km\,s$^{-1}$ \citep{Kharchenko2007}.  The values from the literature are consistent with our derived values but this target remains a potential multiple system due to the high standard deviation. \\

\noindent{\bf CD-363202}: This target only has one observation one from FEROS data, producing a value of 5.2$\pm$0.9\,km\,s$^{-1}$ and one value from the catalogue search, 25.6$\pm$1.0\,km\,s$^{-1}$ \citep{Torres2006}. This is a huge variation in RV; however, it has an extremely high $vsini$ value of 170$\pm$17\,km\,s$^{-1}$\citep{Torres2006}, which severely limits the accuracy of the RV determination.  For this reason it is not considered as a multiple system candidate.\\

\noindent{\bf AK Pic}: There are two observations for this target producing a value of 35.9$\pm$0.9\,km\,s$^{-1}$.  There are six values from VizieR; four of which have associated uncertainties: 32.3$\pm$0.4, 32.1$\pm$0.5, 32.2$\pm$1.5, and 28.1$\pm$2.1\,km\,s$^{-1}$. These are more than 3$\sigma$, $<$33.1\,km\,s$^{-1}$, below our derived value, using their upper uncertainty limits.  Therefore, this target is a potential multiple system.\\

\noindent{\bf BD+012447}:  This target only has one value calculated in this work, 7.7$\pm$0.9\,km\,s$^{-1}$. However, there are seven values from VizieR, two of which have associated uncertainties: 8.3$\pm$0.2 and 9.1$\pm$0.6\,km\,s$^{-1}$.  These values agree with our derived values, and therefore, this target is not considered as a multiple system candidate.\\

\noindent{\bf CD-2711535}: There are two observations for this target; it has a calculated RV value of -6.9$\pm$1.4\,km\,s$^{-1}$.  This standard deviation is well above the average for the sample (0.89\,km\,s$^{-1}$).  There are also two values in VizieR, -6.4$\pm$1.0 and -1.1$\pm$\,km\,s$^{-1}$ from \cite{Torres2006} and \cite{Song2012}, respectively.  The latter value significantly varies from our derived value, and therefore, this target is considered a potential multiple system.\\

\noindent{\bf GJ 3305}: This target has three observations.  It has a calculated RV value of 23.8$\pm$0.5\,km\,s$^{-1}$.There is one value in VizieR 17.6\,km\,s$^{-1}$ from \cite{Reid1995}. However, after private communication with John E. Gizis, it is apparent that the 1$\sigma$ uncertainties are $\approx$17\,km\,s$^{-1}$ for this work, and, therefore, this apparent variation is consistent.  It is not considered as a member of $\beta$-Pic with new RV data from this work.  It is a known extremely eccentric multiple system with an eccentricity of 0.06 and a period of 21.5\,yr \citep{Delorme2012}.\\

\noindent{\bf BD-211074B}: This target has three observations; it has a calculated RV value of 21.6$\pm$0.6\,km\,s$^{-1}$.  There are two values from the VizieR 20.0 and 31.7\,km\,s$^{-1}$ from \cite{Reid1995}.  However, after private communication with John E. Gizis, it is apparent that the 1$\sigma$ uncertainties are $\approx$17\,km\,s$^{-1}$ for this work, and therefore, this apparent variation is consistent.  Therefore, this target is not considered as a multiple system candidate.\\

\noindent{\bf PX Vir}: \cite{Griffin2010} determined the spectroscopic orbit of this system (216.48 day).  In addition to this, we have one value of -13.3$\pm$0.89\,km\,s$^{-1}$, and there are ten values from the VizieR, of which six have associated uncertainties. There are two values from \cite{Maldonado2010} that have a huge variation, which are much higher than the level of uncertainty at -13.1$\pm$0.1 and -2.6$\pm$0.1\,km\,s$^{-1}$, and agrees with the work of \cite{Griffin2010}. Therefore, this target is classified as a multiple system in this work. \\


\noindent{\bf GSC 08077-01788}:  This target has two observations; it has a calculated RV value of 17.6$\pm$0.8\,km\,s$^{-1}$. There are two values from VizieR, \citep{Torres2006}: 24.0$\pm$1.0\,km\,s$^{-1}$ and \citep{Kordopatis2013}: -6.9$\pm$4.1\,km\,s$^{-1}$ . This is a very large variation, and therefore, this target is considered a multiple system in this work. \\

\subsection{Archival multiplicity flags}
\label{sec:Archival}

In addition to querying for any existing RV values, we also queried for multiplicity flags from catalogues using VizieR (\cite{Dommanget2002, Malkov2006, Pourbaix2004, Worley1996}), whether the system has previously been identified to have one or more companions.  Out of our sub-sample of the overall SACY sample and probing only the SB systems, we found the previous multiplicity results: \\

\noindent{\bf HD 33999}: This is a previously known triple-lined system \citep{Dommanget2002} which has been flagged in previous SACY work and identified in this work as well. \\

\noindent{\bf PX Vir}: As discussed in Section~\ref{subsec:archival_rv_variations}, this is a previously identified spectroscopic multiple system which agrees with the variation seen in our catalogue search.

\subsection{Definite SB2 and SB3 candidates}
\label{sec:def_sb2_systems}

Below are any details on the data of postively identified SB2 and SB3 candidates; these targets show a clear double peak in the output of the CCF.  The literature was also queried to use any potential existing information. \\

\noindent{\bf HD 155177:} There are four available data epochs for this object over a wide range of dates (2007-04-19, 2008-10-26, 2012-05-09, and 2012-06-14).  There are two clear peaks that vary across the data epochs indicating strong it is a binary system. No obvious note of spectroscopic multiplicity in the literature.\\

\noindent{\bf HD 33999:} As mentioned in Appendix~\ref{sec:Archival}, this is a previously known triple-lined multiple system.\\

\noindent{\bf HD 309751:} There is only one epoch of data available for this object (2011-11-20); however, the produced CCF profile is extremely clear to identify it as an SB2 system. \cite{Torres2006} identified this as an SB2 system.\\

\noindent{\bf BD-20951:} There are two epochs of data for this target (2012-07-17, 2012-07-31).  There are two clear peaks in the CCFs, indicating strongly it is a bound binary system, . To the best of our knowledge, there is no note of spectroscopic multiplicity in the literature.\\

\noindent{\bf V4046-Sgr:} There are a great many data epochs for this object and it is a well known and studied spectroscopic binary system (e.g., \cite{Quast2000, Argiroffi2012, Donati2011}).\\

\noindent{\bf HD 217379:} There is only one epoch of data for this object (2010-05-25).  Despite this, the CCF provides a clear result that this is a triple-lined system; the signal-to-noise is $\sim$200, and all three peaks are well above the noise level. \cite{Torres2006} identified this as a triple-lined (SB3) spectroscopic system. \\

\noindent{\bf TWA 14:} There are two data epochs available for this object (2012-05-22, 2012-06-04).  The output of the CCF shows two peaks; however the profiles are merged. The two peaks have different depths in each data epoch and appear to switch position between the two epochs, indicating it is a bound binary system. \cite{Weinberger2013} and identified \cite{Jayawardhana2006} this as a double-lined (SB2) system. \\


\noindent{\bf TWA 20:} There are three epochs of data for this target (2012-05-07, 2012-05-22, and 2012-07-03).  There are two strong peaks in all of the CCFs from the RED arm of UVES; the blue is very noisy due to the spectral type of the star, M3.  In addition, \cite{Jayawardhana2006} identify this is a SB2 system, and therefore, it is classified as an SB2 system in this work; however it is not considered a member of TW-Hydrae with new RV values. \\

\noindent{\bf 1RXS J195602.8-320720:}  There is one epoch of data for this target (2011-10-09).  The spectral type of the star is M4, and therefore, the CCF output is quite noisy; however, there is a clear second peak above the level of the noise in both the results from the RED arm of UVES.  To the best of our knowledge, there is no note of spectroscopic multiplicity in the literature. 
\\

\subsection{Questionable SB2 and SB3 candidates}
\label{sub_sec:questionable_sb2}

\setlength{\tabcolsep}{3pt}
\begin{table*}[!tb]
\caption{Details of all initially flagged potential multiple systems.}
\centering
\begin{tabular}{p{2.9cm} p{1.4cm} p{1.7cm} p{1.8 cm} p{1.7cm} p{1.3cm} p{1.5cm} p{1.5cm} p{2.2cm}}
\hline\hline\\[-1ex]
Target & SpT & MJD & RV$_{\mathrm{blue}}$ (km\,s$^{-1}$) & Relative phase ($\phi$)\tablefootmark{a}& Period (day)\tablefootmark{b} & Variability Type\tablefootmark{c} & $\mathrm{v\,sin}\,(i)$ & Association\\\\
\hline\\[-1ex]
CD-423328 & K1 & 56175.398  & 15. 1& - &- & rot & 41.0\tablefootmark{c} & CAR\\
 & & 56179.380  & 15.3 \\[1ex]
CD-433604 & K4 & 54876.233 & 20.4 & 0 & 0.89 & -  & 40.0\tablefootmark{b} & ARG\\
 &  & 56175.381 & 24.1  & 0.717 \\
 &  & 56179.371 & 11.0 & 0.200\\[1ex]
CD-542644 & G5 & 54898.165 & 20.8  & 0 & 1.5 & rot & 39.5\tablefootmark{b} &CAR\\
 &  & 55885.322 & 21.5  & 0.105\\[1ex]
CP-551885\tablefootmark{*} & G5 & 54898.158 & 27.6  & 0 & 0.916 & rot & 44.0\tablefootmark{b} & CAR\\
 &  & 55885.307& 24.0 & 0.674\\[1ex]
GSC 09239-01572 & K7 & 55340.108  & 16.4  & 0 & 1.536 &  - & 41.0\tablefootmark{b} &ECH\\
 &  & 55947.346 & 12.0  & 0.337\\
 &  & 55978.336 & 8.6 & 0.513\\
 &  & 55979.299 & 13.6 & 0.140\\[1ex]
HD 30051\tablefootmark{*} & F2 & 56138.352  & -1.5  & - & - & - & - & THA\\
 &  & 56139.358 & -2.1 \\
 &  & 56140.361  & -2.5 \\[1ex]
TYC 7605-1429-1 & K4  & 55461.359 &  30.7 & - & 0.3514 &  - & - & ABD\\[1ex]
TYC 7627-2190-1 & K2 & 54896.092 & 25.6  & 0 & 0.726 & - & - & ABD\\
 &  & 55416.402 & 27.2 & 0.680\\[1ex]
TYC 8594-58-1 & G8 &  54880.184 & 10.8  & 0 & 0.982 & - & 34.1\tablefootmark{b} & ARG\\
 &  & 56082.022 & 8.2 & 0.868\\
 &  & 56107.997 & 10.9 & 0.319\\[1ex]
\hline
\end{tabular}
\label{table:questionable_sb2}
\tablefoot{\tablefoottext{*}{No longer classified as a member of association with new RV values.}\tablefoottext{a}{Relative phase values (relative to our first observation) were computed using the MJD values from each observation and period values.}  \tablefoottext{b}{\cite{Messina2010, Messina2011}}.   \tablefoottext{c}{\cite{Kiraga2012}: Variability Type: {\bf rot}=rotational variability due to the presence of spots.} }
\end{table*}


%
%
%
%
%
%
%
%

Initially, some targets were flagged as potential multiple systems, as shown in Table~\ref{table:questionable_sb2}.  The targets CD-423328, CD-542644 and CP-551885 have previously been identified as having variable rotational behaviour due to the presence of spots \citep{Kiraga2012}.  However, this does not exclude the possibility that they are multiple systems; therefore, we conduct the same analysis for these targets as for those without any variability flag.

For the targets shown in Table~\ref{table:questionable_sb2}, we conducted the following analysis to determine whether the deformity could be the solely the result of spots on the surface of the star.


The intensity of Ca II H (3968.5~$\AA$) and Ca II K (3933.7~$\AA$) emission respond to the amount of non-thermal heating of the chromosphere, which can be caused by areas of concentrated magnetic field \citep{Leighton1959}; therefore, there is a strong link between spot coverage and activity.  We can use this to our advantage in our CCF output to differentiate between spots and merged-SB2 components.

We fitted one Gaussian profile to the CCF output, assuming it was a single star, producing a system velocity ($v_{\mathrm{sys}}$) and then looked for the specific velocities of the deformities in the profile ($v_{\mathrm{d1}}, v_{d2}$, etc.).  We then compared the system velocity and deformity velocity values to the line profiles of Ca II H and K; the chromosphere can emit isotropically at the sytem velocity and have components of emission from the strong areas of magnetism, corresponding to the deformity velocities.  If the Ca II H and K lines had a multiple-component structure with peaks corresponding to the values of $v_{\mathrm{sys}}$ and $v_{\mathrm{d1}}, v_{\mathrm{d2}}$ etc. then it is very likely that the candidate multiple system is a single system with strong activity.  If this was not the case but the profiles of Ca II H and K had a multiple component structure, we also checked whether the multiple components had peak values corresponding to a multiple-line fit to the CCF, which assume the system is multiple.


The graphs presented in this section are in the following format.  There are four panels for each UVES observation from left to right (1-4). 

\begin{itemize}
\item {\it Panel 1:} The CCF output for the BLUE chip of UVES (blue line) and single Gaussian fit (black line).  Phase values, $\phi$, are shown from Table~\ref{table:questionable_sb2} where available. 
\vspace{0.1cm}
\item {\it Panel 2:} The residuals from each of the fits, all graphs have the same y-axis range for easy comparison of the goodness of fit. The parameter $\sigma$ represents the average standard deviation in the residuals from the fit.  
\vspace{0.1cm}
\item {\it Panel 3:}  The normalised flux of chromospheric line Ca II H, 3968.5~$\AA$, in velocity space (black line); overplotted is the normalised and reversed CCF profile from {\it Panel 1} (blue dotted line) as well as $v_{\mathrm{sys}}$ and $v_{\mathrm{d1, 2}}$ (red vertical lines).  
\vspace{0.1cm}
\item {\it Panel 4} has the same format are {\it Panel 3} but is at 3933.7~$\AA$, for the chromospheric line Ca II K. 
\end{itemize}

\noindent{\bf TYC 7605-1429-1}: This target was initially identified as a binary candidate due to its abnormal CCF profile. However, TYC 7605-1429-1 only has one epoch of data, which is not conclusive from this analysis, as shown in Figure~\ref{fig:TYC 7605-1429-1}.  It has an extremely broad component with a FWHM of $\approx$62\,km\,s$^{-1}$, making the RV determination difficult. \cite{Torres2006} quote a value of 28.6\,km\,s$^{-1}$ which agrees with our derived value of 31.2$\pm$4.0\,km\,s$^{-1}$. \cite{Messina2010} flagged this target as a single system. \\

\begin{figure}[h]
\includegraphics[width=0.49\textwidth]{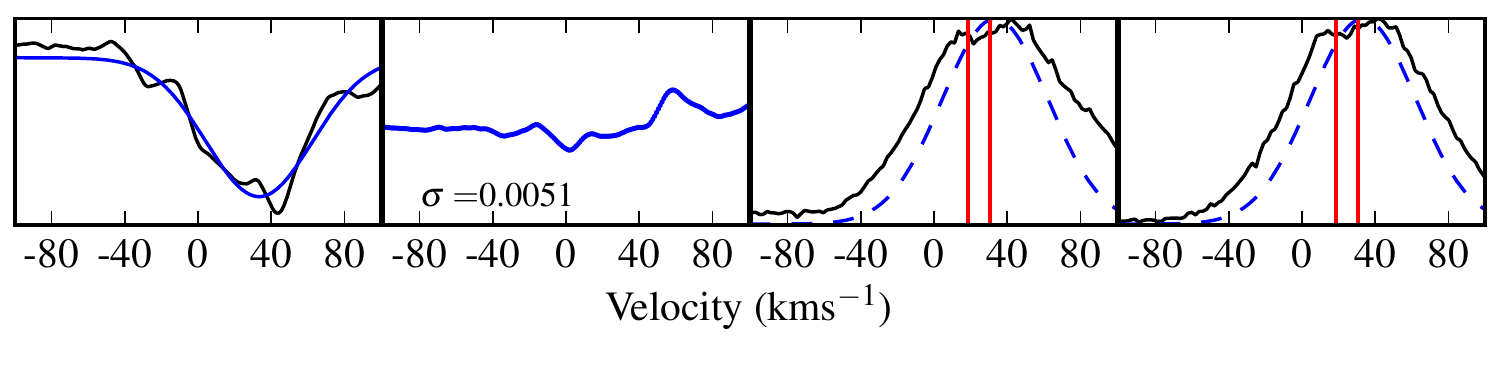}
\vspace{-0.5cm}
\caption{TYC 7605-1429-1: 2012-09-22.}
\label{fig:TYC 7605-1429-1}
\end{figure}

\noindent{\bf HD 30051}: There are three available epochs for this target, 2012-07-30, 2012-07-31, and 2012-08-01.  This is an F2-type star, and therefore, does not have strong chromopsheric activity, as shown in the lack of emission in Panels 3 and 4 of Figure~\ref{fig:HD30051}.  This star was initially classified as a multiple system due to the two component structure of the CCF profile, which potentially consists of a rapid rotator, producing the broad component with a slower-rotating companion, narrow component.  However, there is no significant variation in system velocity (-1.5, -2.1 and, -2.5) between observations, making it extremely unlikely to be a close binary system with period $>$ 1 day. \\

\begin{figure}[h]

\includegraphics[width=0.49\textwidth]{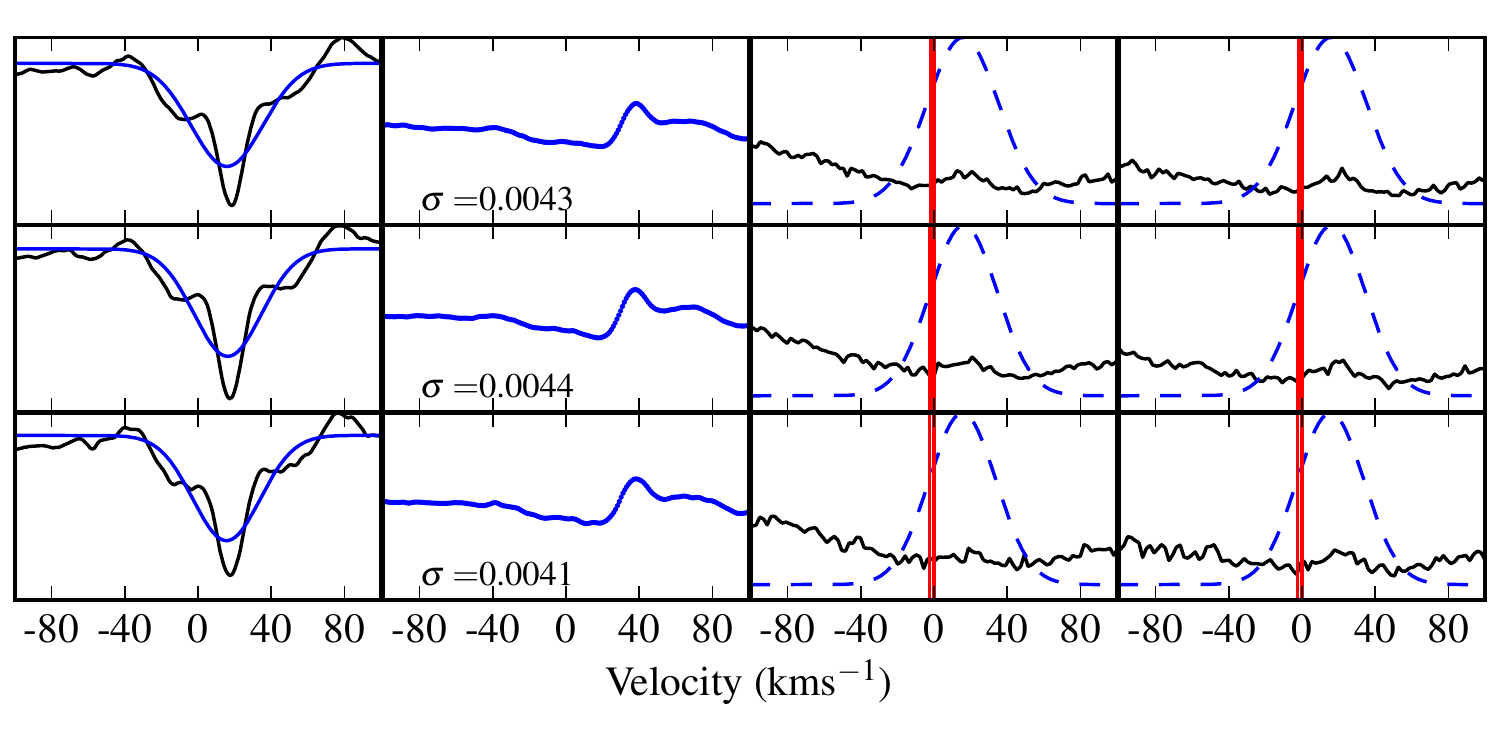}
\vspace{-0.5cm}
\caption{HD 30051: 2012-07-30, 2012-07-31, and 2012-08-01.}
\label{fig:HD30051}
\end{figure}

\noindent{\bf CD-542644}:  There are two epochs for this target 2009-03-08 and 2011-11-20, as seen in Figure~\ref{fig:CD-542644}.  The velocities for $v_{\mathrm{sys}}$ and $v_{\mathrm{d1}}$ do not correspond well to the shape of the line profiles.  The CCF profiles are also quite complex, making it hard to determine the correct velocities.  Based on this analysis, we do not consider this target a multiple system.  \\

\begin{figure}[h]

\includegraphics[width=0.49\textwidth]{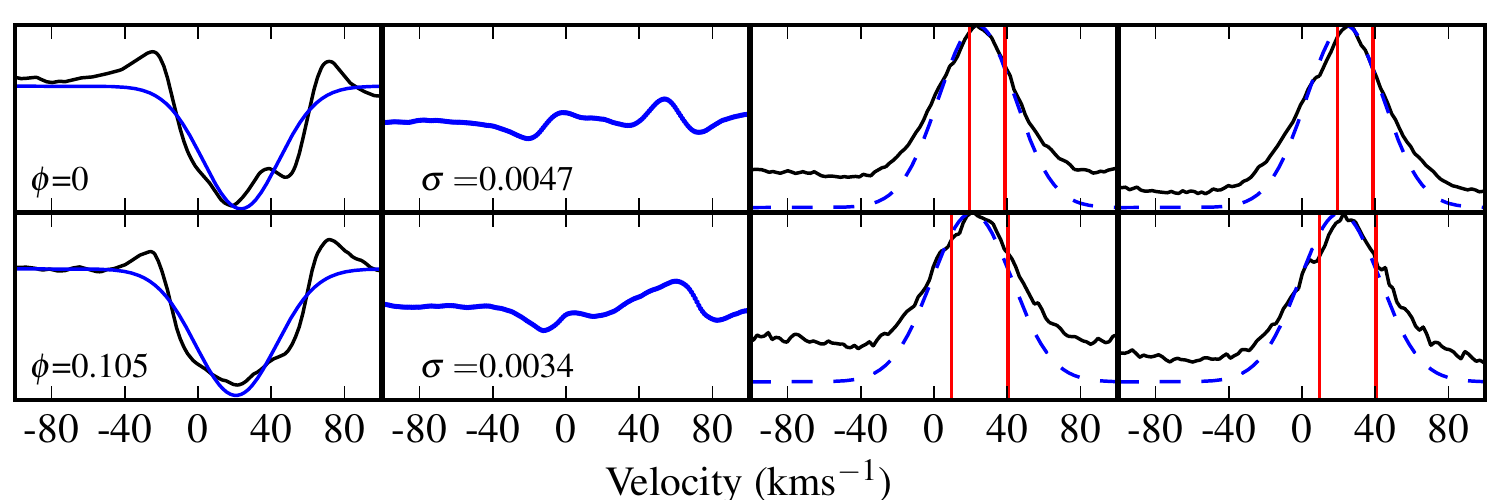}
\vspace{-0.5cm}
\caption{CD-542644: 2009-03-08, and 2011-11-20.}
\label{fig:CD-542644}
\end{figure}

\noindent{\bf CP-551885}:  There are two epochs for this target 2009-03-08 and 2011-11-20, as seen in  Figure~\ref{fig:CP-551885}.  The line profiles displayed in panels 3 and 4 show a clear two component structure.  However, the picture we have is confusing; the $v_{\mathrm{sys}}$ value agrees well in both epochs; however, $v_{\mathrm{d1}}$ does not.  Instead, it is one of the velocities fitted using two Gaussians that matches with the secondary peak $\approx$5\,km\,s$^{-1}$. In addition to this, if we look at the $v_{\mathrm{sys}}$ values, there is no considerable variation: 27.6 ($\phi=0$), 21.5 ($\phi=0.674$) and \citep{Torres2006} 23.1\,km\,s$^{-1}$ from previous work, assuming the period of rotation is synchronised with the orbital period. Based on these arguments, we do not consider this target a multiple system. \\

\begin{figure}[h]

\includegraphics[width=0.49\textwidth]{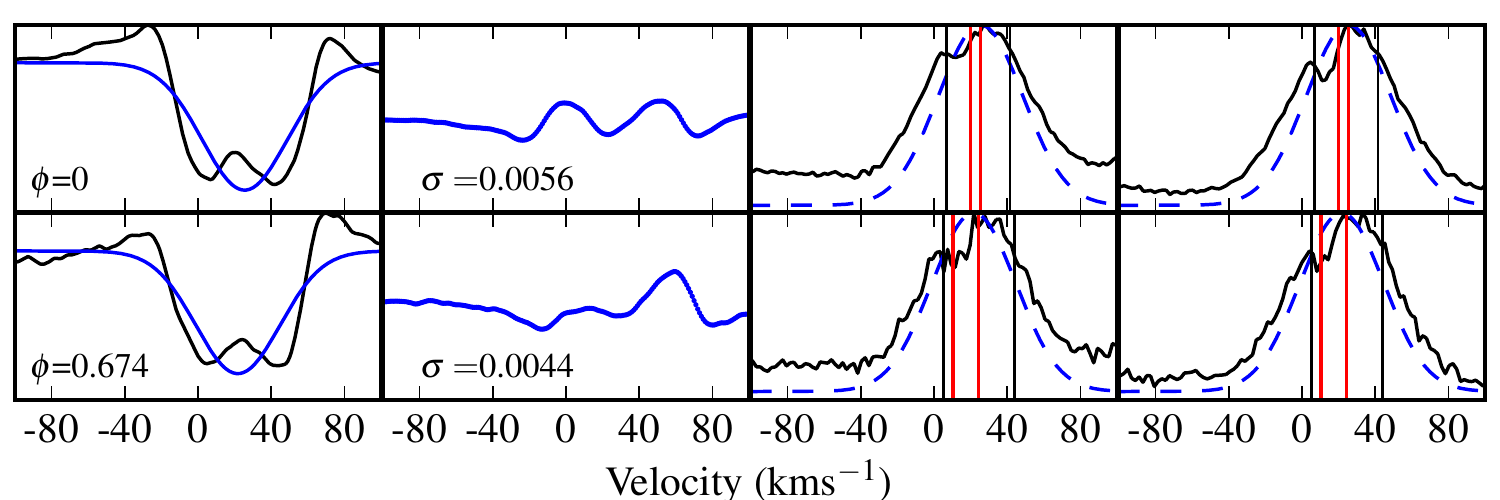}
\vspace{-0.5cm}
\caption{CP-551885: 2009-03-08, and 2011-11-20.}
\label{fig:CP-551885}
\end{figure}

\noindent{\bf CD-423328}:  There are two epochs for this target 2012-09-05 and 2012-09-09, as seen in Figure~\ref{fig:CD-423328}.  The lines from a two-component Gaussian fit agree with the peaks seen in the profiles of Ca II H and K.  \cite{Torres2006} also quote a value of 24.6$\pm$1.0\,km\,s$^{-1}$, which is hugely different from our value derived from a single Gaussian fit (15.2$\pm$0.1\,km\,s$^{-1}$).  \cite{Kiraga2012} classified this object as rotationally variable due to the presence of spots; however, this does not exclude the possibility that it is a binary system.  Due to this line-profile analysis and the variation in RV value, this target is classified as a potential multiple system.   \\

\begin{figure}[h]

\includegraphics[width=0.49\textwidth]{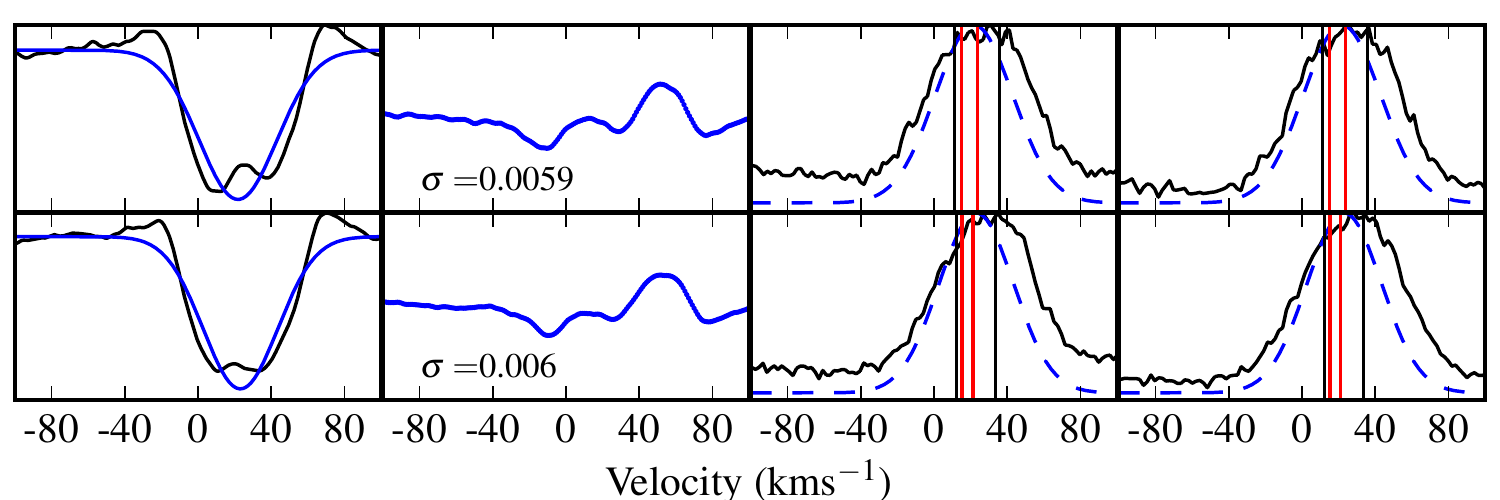}
\vspace{-0.5cm}
\caption{CD-423328: 2012-09-05, and 2012-09-09.}
\label{fig:CD-423328}
\end{figure}

\noindent{\bf CD-433604}:  There are three available epochs for this target, 2009-02-14, 2012-09-05, and 2012-09-09, as seen in Figure~\ref{fig:CD-433604}.  It is extremely unlikely that it is an SB2 system, as the flux ratio of the two components is variable epoch to epoch (0.97, 0.64, 0.56 respectively). The line profile analysis is inconclusive; epochs 2012-09-05 and 2012-09-09 show a two-component structure in Ca II H and K; however, the derived $v_{\mathrm{sys}}$ and $v_{\mathrm{d1}}$ values do not correspond to the peaks of these components. The calculated RV values (20.4, 24.1, and 11.0) are extremely variable, but the fits are poor, as seen in panel 2 of Figure~\ref{fig:CD-433604}.  There is one other existing value (priv. communication, C. A. O. Torres) of 21.4, and therefore, if one considers three of the four values to compute an average, there is no significant variation (22.0$\pm1.9$\,kms$^{-1}$).  At this time, the evidence for multiplicity is not strong enough for us to consider this target as a multiple system.\\



\begin{figure}[h]
\begin{center}
\includegraphics[width=0.49\textwidth]{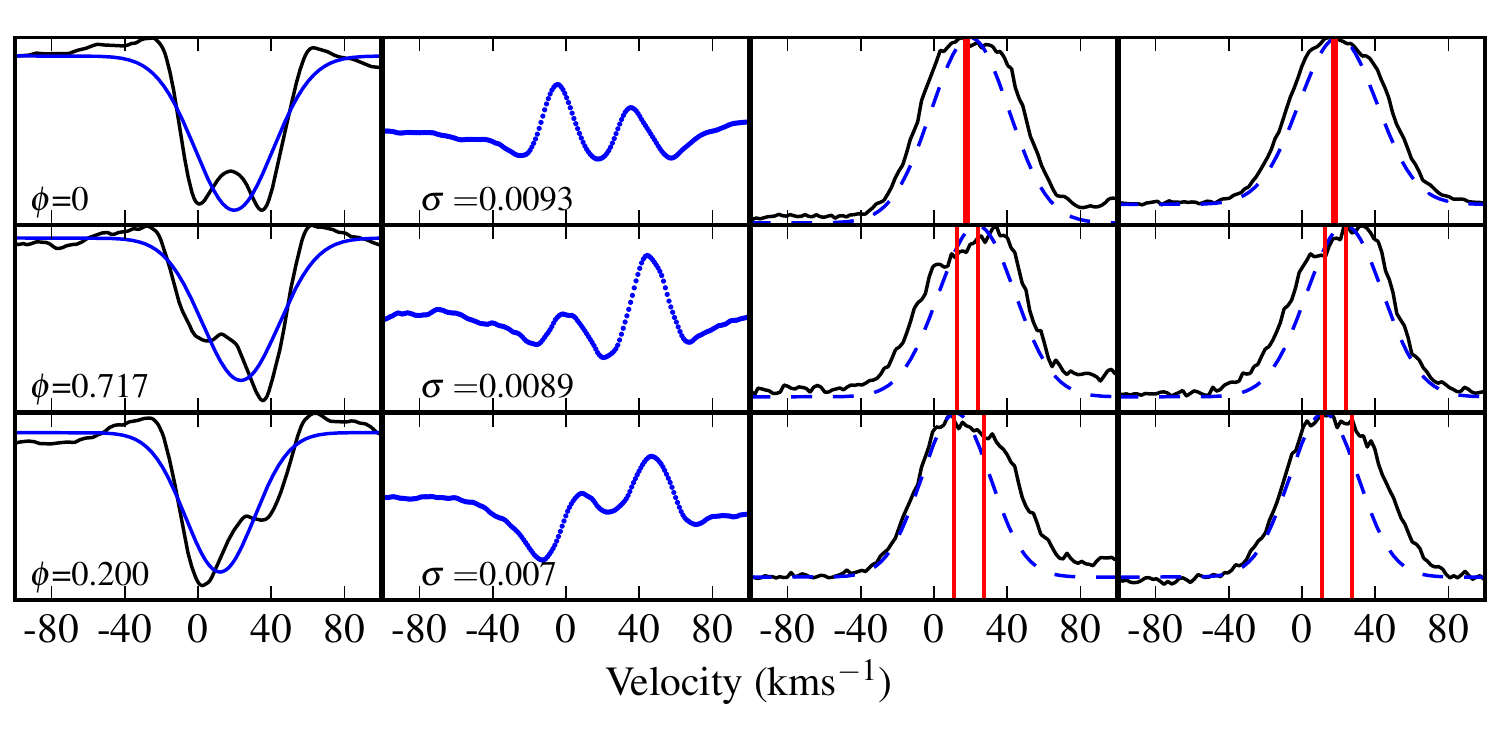}
\end{center}
\vspace{-0.5cm}
\caption{CD-433604: 2009-02-14, 2012-09-05, and 2012-09-09.}
\label{fig:CD-433604}
\end{figure}

\noindent{\bf GSC 09239-01572}: There are four epochs available for this object, 2010-05-24, 2012-01-21, 2012-02-21, and 2012-02-22.  In Figure~\ref{fig:GSC 09239-01572}, one can see the agreement between the deformity velocities and the line profiles of Ca II H and K is extremely good for all four epochs.  On this basis, we conclude that the CCF profile is the result of star spots as opposed to multiplicity.\\

\begin{figure}[h]
\includegraphics[width=0.49\textwidth]{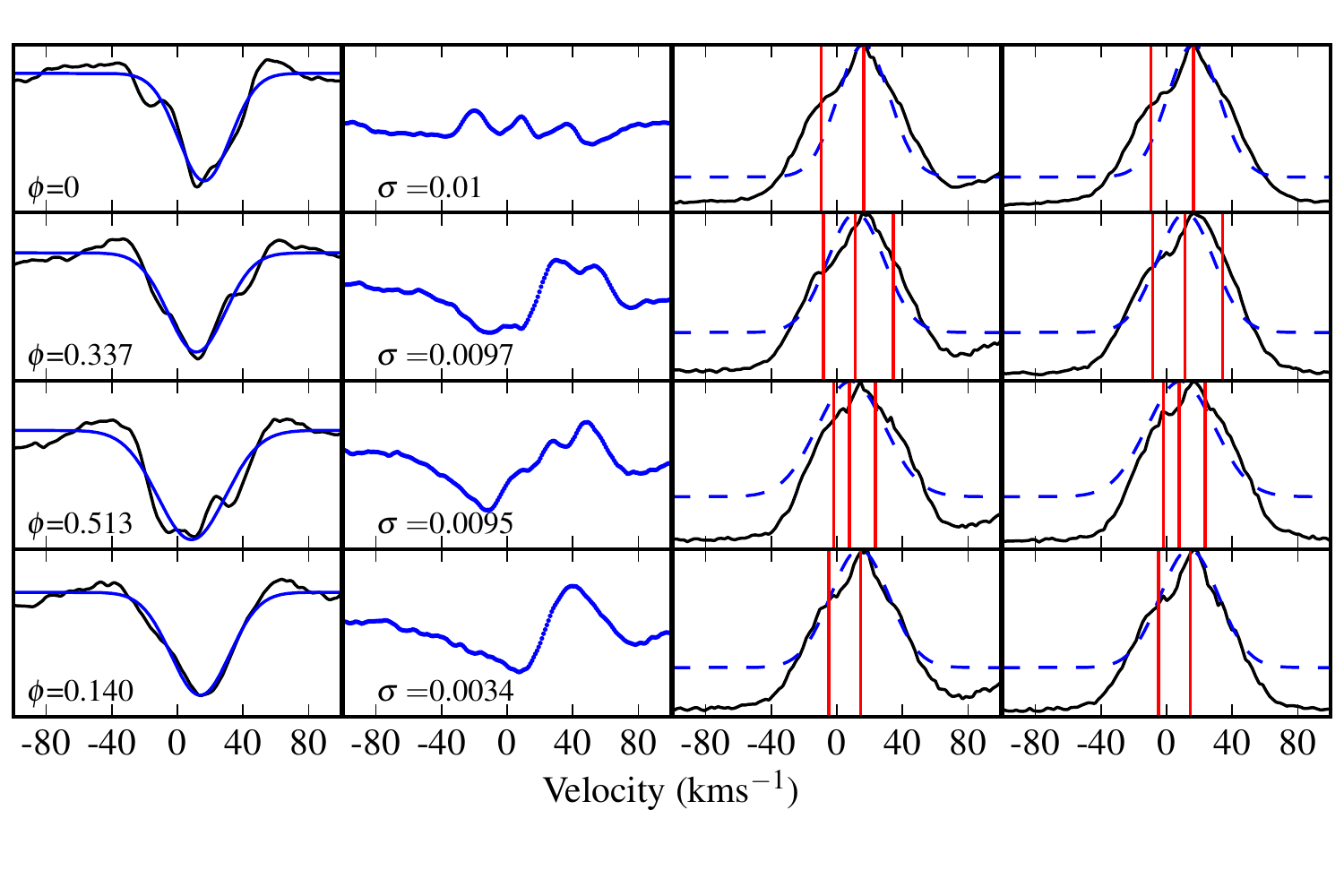}
\vspace{-0.5cm}
\caption{GSC 09239-01572: 2010-05-24, 2012-01-21, 2012-02-21, and 2012-02-22.}
\label{fig:GSC 09239-01572}
\end{figure}

\noindent{\bf TYC 7627-2190-1}: There are two epochs available for this object: 2009-03-06 and 2010-08-08, as seen Figure~\ref{fig:TYC 7627-2190-1}.  From the fitting of a single system velocity to the profiles there is no significant variation betwen the observations (25.6 and 27.2).  The value of the system velocity is extremely close to that of the deformity velocity; this makes it extremely difficult to see the separate components in this analysis.  However, we only have two epochs, and therefore based on this evidence, we do not classify this as a multiple system. \\

\begin{figure}[h]
\begin{center}
\includegraphics[width=0.49\textwidth]{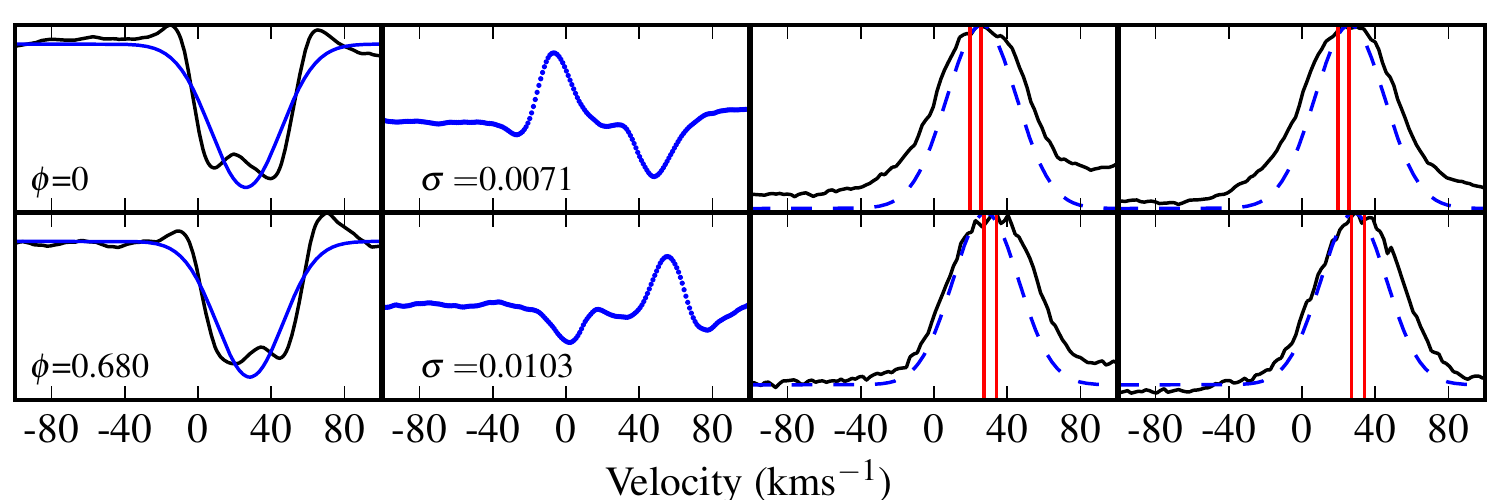}
\end{center}
\vspace{-0.5cm}
\caption{TYC 7627-2190-1: 2009-03-06, and 2010-08-08.}
\label{fig:TYC 7627-2190-1}
\end{figure}

\noindent{\bf TYC 8594-58-1}: There are three epochs of data for this target: 2009-02-18, 2012-06-04 and 2012-06-29.  However, much like TYC7627-2190, the value of the deformity velocity is extremely close to the system velocity, and therefore, seeing the two components separately is extremely difficult.  There is no significant variation in the system velocity between all three epochs (10.8, 8.2, and 10.9, respectively). Figure~\ref{fig:TYC 8594-58-1} also shows the velocities when fitting the CCF with two Gaussians, as shown as black vertical lines.  Neither of these techniques provide strong evidence for or against the multiplicity of the system.  Therefore, as it statistically much more likely to be a single system, it is not considered a multiple system in this work.

\begin{figure}[h]
\begin{center}
\includegraphics[width=0.49\textwidth]{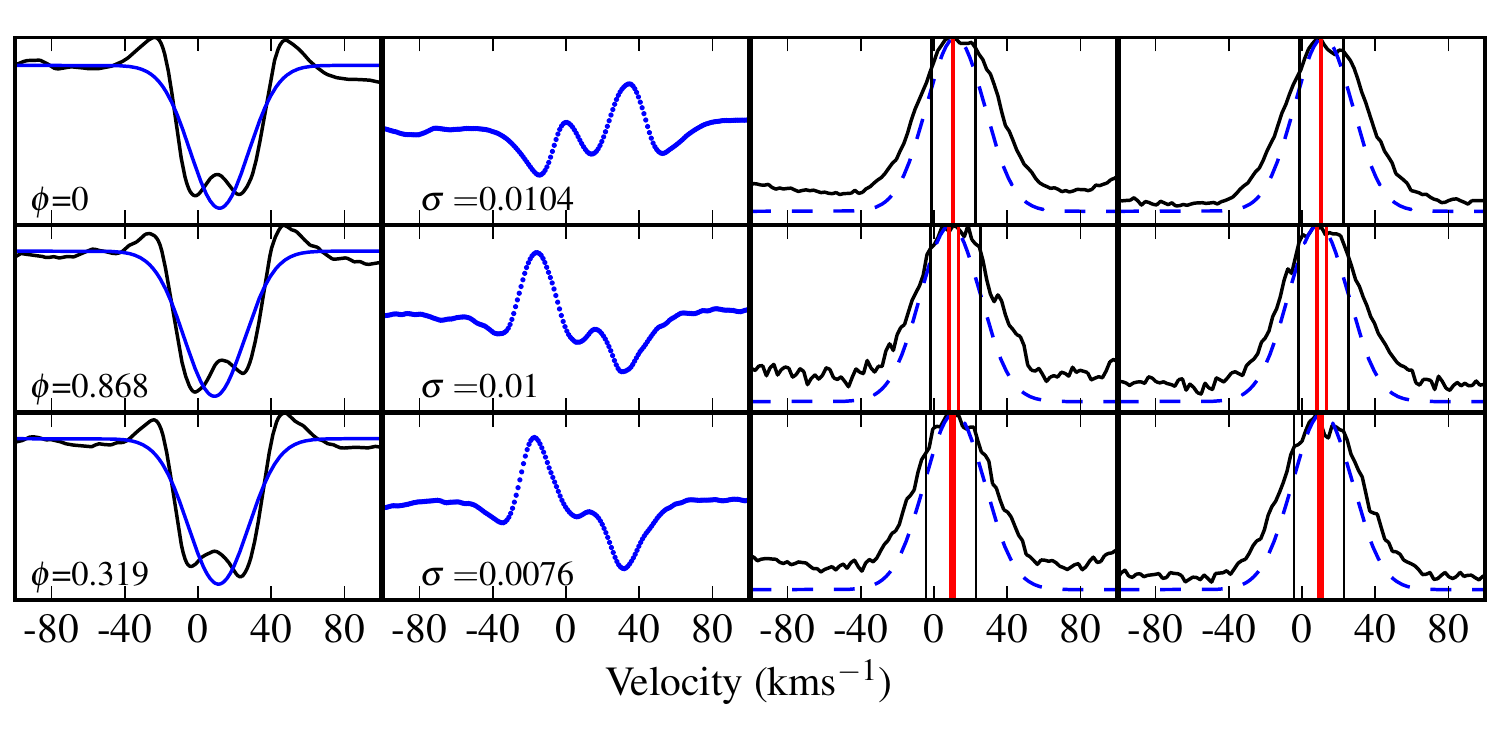}
\end{center}
\vspace{-0.5cm}
\caption{TYC 8594-58-1: 2009-02-18, 2012-06-04, and 2012-06-29.}
\label{fig:TYC 8594-58-1}
\end{figure}


\newpage
\bibliography{biblio1}


\Online

\onecolumn
\section{Radial velocity values}
\label{appendix:vrad}



\tablefoottext{*}{No longer classified as a member of association with new RV values.}\\


\noindent\tablefoottext{a}{{RV values from a single-component Gaussian fit; see Section~\ref{sub_sec:questionable_sb2} for details on target.}}\\

\noindent\tablefoottext{1}{Uncertainties for UVES and FEROS data: If more than one observation was available uncertainties were taken to be the standard deviation from the mean value for each target. If there was only one observation the uncertainty has a value of 0.89\,km\,s$^{-1}$, one standard deviation of the entire sample.}\\

\noindent\tablefoottext{2}{Uncertainties from \cite{deBruijne2012} have values $(A)$: the standard errors are generally reliable or $(B)$: potential, small, uncorrected, systematic errors.}\\

\noindent\tablefoottext{3}{\noindent{{\bf References:} a: I/196 \citep{Turon1993}, b: I/306A/ \citep{Ivanov2008}, c:III/105/ \citep{Andersen85}, d: III/198/ \citep{Reid1995}, e: III/249/ \citep{Malaroda2006}, f: III/252 \citep{Gontcharov2006}, g: III/254/ \citep{Kharchenko2007}, h: III/265/ \citep{Siebert2011}, i :J/A+A/423/517/ \citep{Rocha2004}, j: J/A+A/460/695/ \citep{Torres2006}, k: J/A+A/521/A12/ \citep{Maldonado2010}, l: J/A+A/530/A138/ \citep{Casagrande2011}, m: J/A+A/531/A8/ \citep{Jenkins2011}, n: J/A+A/546/A61/ \citep{deBruijne2012}, o: J/A+A/551/A46/ \citep{Martin2013}, p: J/A+AS/142/275/ \citep{Strassmeier2000}, q: J/AJ/133/2524/ \citep{White2007}, r: J/AJ/144/8/ \citep{Song2012}, s: J/AZh/83/821/ \citep{Bobylev2006}, t: J/AcA/62/67/ \citep{Kiraga2012}, u: V/137D/XHIP \citep{Anderson2012}, v: III/272/ \citep{Kordopatis2013}.}}

\newpage

\begin{landscape}
\begin{table}
\caption{Summary table of UVES radial velocity values for SB2 and SB3 systems.}
\label{table:sb2_vr_values}

\end{table}
\noindent\tablefoottext{*}{No longer classified as a member of association with new RV values.}\\
\tablefoottext{a}{From CCF profiles in both epochs, it is difficult to accurately compute the flux ratio; however, the two components are very similar, and thus, we assign a value of 1.00.}
\end{landscape}

\clearpage

\noindent\tablefoottext{*}{No longer classified as a member of association with new RV values.}\\
\tablefoottext{a}{RV values from a single-component Gaussian fit; see Section~\ref{sub_sec:questionable_sb2} for details on target.}\\


\onecolumn
\newpage
\section{List of SACY association members}
\label{appendix:members}

\begin{longtable}{p{3.7cm} p{1.8cm} p{1.8cm} p{1.3cm} p{1.3cm} p{1.4cm} p{1.4cm} p{1.4cm} p{1.4cm}}
\caption{Full member list for the SACY associations used in this work.} \label{tab:member_list}\\
\hline\hline\\
  \multicolumn{1}{l}{ID} &
  \multicolumn{1}{l}{RA} &
  \multicolumn{1}{l}{DEC} &
  \multicolumn{1}{l}{V Mag.} &
  \multicolumn{1}{l}{X} &
  \multicolumn{1}{l}{Y} &
  \multicolumn{1}{l}{Z} &
  \multicolumn{1}{l}{D} &
  \multicolumn{1}{l}{SpT} \\\\
\hline\\
\endfirsthead

\hline\hline\\
  \multicolumn{1}{l}{ID} &
  \multicolumn{1}{l}{RA} &
  \multicolumn{1}{l}{DEC} &
  \multicolumn{1}{l}{V Mag.} &
  \multicolumn{1}{l}{X} &
  \multicolumn{1}{l}{Y} &
  \multicolumn{1}{l}{Z} &
  \multicolumn{1}{l}{D} &
  \multicolumn{1}{l}{SpT} \\\\
\hline\\
\endhead
\multicolumn{9}{c}{AB Doradus members} \\\\
\hline\\
  HD 36705   & 05 28 44.4 & -65 26 47 & 13.2   & 1.2 & -12.7 & -8.3 & 15.2    & M4              \\
  AB Dor       & 05 28 44.8 & -65 26 55 & 6.88   & 1.2 & -12.7 & -8.3 & 15.2    & K0              \\
  AK Pic      & 06 38 00.4 & -61 32 00 & 6.27   & 0.4 & -19.2 & -9.1 & 21.3    & G2              \\
  BD+012447   & 10 28 55.5 & +00 50 28 & 9.65   & -2.0 & -4.1 & 4.9 & 6.7    & M1              \\
  V429 Gem     & 07 23 43.6 & +20 24 59 & 10.00  & 23.8 & -7.7 & 7.2 & 26.0     & K5              \\
\end{longtable}
\tablefoot{\tablefoottext{*}{No longer classified as a member of association with new RV values.}
\tablefoottext{a}{No SIMBAD entry within 5 arcseconds.}}

\end{appendix}




\end{document}